\documentclass[11pt]{iopart}
\usepackage{graphicx}
\usepackage{subfigure}
\usepackage{morefloats}
\usepackage{color}
\usepackage{setspace}
\usepackage{floatrow}
\usepackage{bm}
\usepackage{color}
\begin{document}

\title[Filament motion in a realistic tokamak geometry]{\textbf{Numerical investigation of isolated filament motion in a realistic tokamak geometry}}
\author{ N. R. Walkden$^{1,2}$, B. D. Dudson$^{2}$, L. Easy$^{1,2}$, G. Fishpool$^{1}$ and J. T. Omotani$^{1}$
        \\ \small{$^{1}$ CCFE, Culham Science Centre, Abingdon, OX14 3DB, UK} 
        \\ \small{$^{2}$ York Plasma Institute, Department of Physics, University of York, Heslington, York, YO10 5DD, UK} 
        \\ Email: \texttt{nick.walkden@ccfe.ac.uk} }
\date{}

\begin{abstract}
This paper presents a numerical investigation of isolated filament dynamics in a simulation geometry representative of the scrape-off layer (SOL) of the Mega Amp Spherical Tokamak (MAST) previously studied in [N.R.Walkden \emph{et.al}, Plasma Phys. Control. Fusion, \textbf{55} (2013) 105005]. This paper focuses on the evolution of filament cross-sections at the outboard midplane and investigates the scaling of the centre of mass velocity of the filament cross-section with filament width and electron temperature. 
\\By decoupling the vorticity equation into even and odd parity components about the centre of the filament in the bi-normal direction parallel density gradients are shown to drive large velocities in the bi-normal (approximately poloidal) direction which scale linearly with electron temperature. In this respect increasing the electron temperature causes a departure of the filament dynamics from 2D behaviours. 
\\Despite the strong impact of 3D effects the radial motion of the filament is shown to be relatively well predicted by 2D scalings. The radial velocity is found to scale positively with both electron temperature and cross-sectional width, suggesting an inertially limited nature. Comparison with the two-region model [J. R. Myra \emph{et.al}, Phys. Plasmas, \textbf{13} (2006) 112502]  achieves reasonable agreement when using a corrected parallel connection length due to the neglect of diamagnetic currents driven in the divertor region of the filament.  
\\Analysis of the transport of particles due to the motion of the filament shows that the background temperature has a weak overall effect on the radial particle flux whilst the filament width has a strong effect. 
\end{abstract}

\section{Introduction}
Controlling particle and heat fluxes to material surfaces is an essential step towards achieving net energy gain from a high powered fusion device. Excess heat fluxes to divertor surfaces will result in melting damage requiring repair and limiting operation. Excess particle fluxes can cause erosion of materials, again causing damage but also providing an influx of impurities into the plasma thereby degrading its performance. In the last decade it has become clear that particle transport into the scrape-off layer (SOL) cannot be accurately captured by local transport models such as the diffusion-advection paradigm \cite{NaulinJNM2007}. Instead over $50$\% of particle transport can be mediated by the propagation of large, coherent, field aligned structures called filaments (also commonly referred to as blobs) \cite{BoedoPoP2003}. Filaments have been observed on many tokamaks 
\cite{BoedoPoP2001,MaquedaRSI2001,GravesPPCF2005,CarrerasPoP2001,LiuJNM2011,XuNF2009,KirkPPCF2006,DudsonPPCF2008,AyedPPCF2009} and other magnetically confined plasma devices \cite{HappelPRL2009,MullerPoP2007,AntarPRL2001} suggesting a rather universal nature. Recently the formation of a density shoulder in the SOL density profile \cite{AsakuraJNM97,UmanksyPoP98,LaBombardNF2000,LaBombardPoP2001} has been linked to an increase in filamentary transport \cite{MilitelloArXiV2015} as the gas fuelling rate of the plasma is increased. This may be favourable as it spreads the impact of erosion over a larger area on the divertor surface. It runs the risk however of causing excess interaction with first-wall materials, leading to damage of the first wall and dilution of the core due to an influx of first-wall impurities \cite{HendersonNF2014}. Predictive control of these effects is desirable, however before any level of prediction can be made, a fundamental understanding of the processes governing filament motion must first be established.
\\ \\Models of filament motion have been traditionally based on 2D interchange motion \cite{KrashenninikovPLA2001,YuPoP2003,D'IppolitoCPP2004,GarciaPoP2005,YuPoP2006,OmotaniArXiV2015} where variation in the magnetic field structure leads to a  non-zero divergence of the diamagnetic drift of ions and electrons. This arises due to the inhomogeneity of the background magnetic field and drives a diamagnetic current across the filament. This drive is balanced by dissipation either due to aerodynamics drag, termed the inertially limited regime, or by sheath currents, termed the sheath limited regime. In each case the result is that the filament propagates radially due to an $\textbf{E}\times\textbf{B}$ velocity as a result of a dipole potential forming across its cross-section. The formation of the potential dipole has been observed experimentally \cite{KatzPRL2008,TheilerPoP2011}. For an excellent review of 2D models of filament motion and their comparison with experiment see ref \cite{D'IppolitoReview}. In recent years models for filament motion have been extended to 3D in a basic slab geometry \cite{AngusPRL2012,AngusPoP2012,EasyPoP2014} and in more realistic geometries \cite{WalkdenPPCF2013,HalpernPoP2014,WalkdenThesis}. This paper follows from this work and investigates in detail the motion of filaments at the outboard midplane and how changes in temperature and filament width affect the resulting transport of particles. 
\\ \\This paper is organized as follows: Section \ref{Sec:Model} introduces the model used for filament simulations and describes the role played by 3D effects. Section \ref{Sec:Geom} describes briefly the magnetic geometry employed to simulate the MAST SOL, as well as providing details of the numerical methods used in BOUT++ \cite{BOUT++Paper}. Section \ref{Sec:Dyn} presents the dynamics of filaments across a temperature and filament width scan at the outboard midplane and compares the results to the two-region model \cite{MyraPoP2006,RussellPoP2007}. Section \ref{Sec:Transp} gives a detailed analysis of particle transport due to filament motion and assesses dominant factors. Finally \ref{Sec:Conc} discusses the validity and implications of the results and concludes. 
 
\section{Model}
\label{Sec:Model}
\begin{figure}[htbp]
\centering
\includegraphics[width=\textwidth]{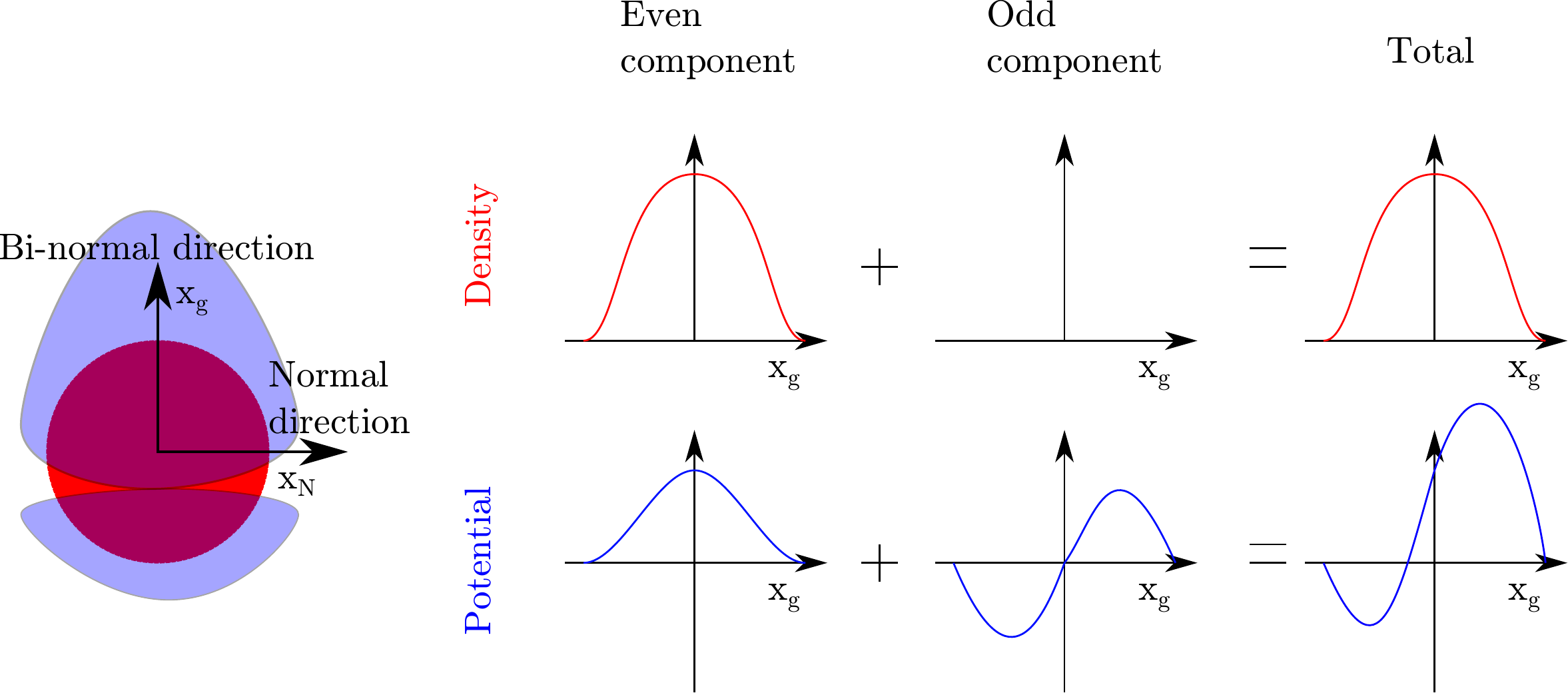}
\caption{Schematic illustration of the deconstruction of the potential and density profiles into odd and even functions about the filament centre in the bi-normal direction. The density is assumed to be entirely even whilst the potential is a superposition of the two components.}
\label{Fig:Sym_schem}
\end{figure}

The model used in this paper to investigate the dynamics of filaments is based on the continuity of density and current. The model follows the drift ordering \cite{Simakov-Catto-Eqns} and additionally makes the Boussinesq approximation and assumes that the plasma is isothermal and electrostatic. This prevents the model from capturing effects due to magnetic or thermal fluctuations. The low pressures in L-mode filaments lead to low values of $\beta$, however high pressure gradients may still make the electrostatic assumption fragile \cite{RogersPRL98,LaBombardPoP2008}. Since this work seeks to establish the connection between basic models of filament motion and 3D models these effects have been neglected, however their inclusion should be treated as a priority for future study. Since thermal conduction time scales are much faster than particle loss time scales (typically $L_{||}/c_{s}$), temperature fluctuations are likely to be drained to the divertor much more quickly than density fluctuations, so the isothermal assumption may be justified in the far SOL. This is a fragile assumption however and non-isothermal effects will be considered in a future paper. For the present paper it is instructive to use a highly reduced a model to isolate the causes of any effects in the filament dynamics observed. In this vein ions are assumed cold such that $T_{i} \ll T_{e}$ which eliminates FLR effects associated with the diamagnetic contribution to vorticity \cite{BisaiPoP2013,ManzPoP2013}. Finally the effects of parallel ion streaming are neglected as slow in comparison to the cross-field evolution of the filament and electron inertia is neglected. These latter two assumptions have recently been shown to be justified in 3D slab simulations \cite{EasyPoP2014}. Under this set of assumptions the governing equations of the system are:
\begin{equation}
\label{Eqn:Dens}
\frac{\partial n}{\partial t} + \nabla\cdot\left(n\textbf{v}\right) = 0 \quad \Rightarrow \quad \frac{dn}{dt} = 2c_{s}\rho_{s}\bm{\xi}\cdot\nabla n + \frac{1}{e}\nabla_{||}J_{||}
\end{equation}
\begin{equation}
\label{Eqn:Vort}
\nabla\cdot\textbf{J} = 0 \quad \Rightarrow \quad \rho_{s}^{2}n\frac{d\nabla_{\perp}^{2}\phi}{dt} = 2c_{s}\rho_{s}\bm{\xi}\cdot\nabla n + \frac{1}{e}\nabla_{||}J_{||}
\end{equation}
with 
\begin{equation}
\label{Eqn:Jpar}
\eta_{||}J_{||} = T_{e}\nabla_{||}\left(\ln(n) - \phi\right)
\end{equation}
\begin{equation}
\frac{d}{dt} = \frac{\partial}{\partial t} + \textbf{v}_{E}\cdot\nabla
\end{equation}
and
\begin{equation}
\textbf{v}_{E} = c_{s}\rho_{s}\textbf{b}\times\nabla\phi
\end{equation}
where $n$ is the electron density (ion density is inferred through quasineutrality), $\textbf{J}$ is the total current density, $J_{||}$ is the component of current density parallel to the magnetic field and $\phi = \Phi/T_{e}$ is the potential normalized to the electron temperature $T_{e}$ in eV. $\bm{\xi} = \left(\nabla\times\left(\textbf{b}/B\right)\right)B \approx \textbf{b}\times\bm{\kappa}$ is the polarization vector arrising from the divergence of the magnetization drift where $\textbf{b}$ is the magnetic field tangency vector and $\bm{\kappa} = \textbf{b}\cdot\nabla\textbf{b} = -\textbf{b}\times\nabla\times\textbf{b}$ is the magnetic field curvature vector. $c_{s} = \sqrt{eT_{e}/m_{i}}$ is the Bohm sound speed, $\rho_{s} = c_{s}/\Omega_{i}$ is the Bohm gyro-radius and $\Omega_{i} = eB/m_{i}$ is the ion gyro-frequency. $\eta_{||}$ is the parallel resistivity here taken as the Spitzer resistivity, $\eta_{||} = m_{e}/(1.96ne^{2}\tau_{e})$ with the electron ion collision time given by $\tau_{e} = 3\sqrt{m_{e}}T_{e}^{3/2}/(4\sqrt{2\pi}n\lambda e^{4}) = 3.44\times 10^{11}\left(\frac{T_{e}}{eV}\right)^{3/2}\left(\frac{n}{m^{-3}}\right)^{-1}\lambda^{-1}$ where $\lambda\sim10$ is the Coulomb logarithm. Terms on the RHS of equation \ref{Eqn:Dens} are smaller than leading order, however Angus \emph{et.al} have shown that they must be retained to properly capture the $k_{\perp}$ dependence in the linear regime of the resistive drift wave instability \cite{AngusPRL2012,AngusPoP2012}. 
\\Three-dimensional effects are incorporated into the system through parallel Ohm's Law, equation \ref{Eqn:Jpar}. The inclusion of these effects introduces a non-dipolar component of the electrostatic potential, thereby inducing motion in the poloidal direction alongside the filament's radial motion \cite{AngusPoP2012,WalkdenPPCF2013,EasyPoP2014}. To illustrate this the symmetry properties of the filament potential can be considered. In the idealized situation, often studied in two and three dimensions, the density is taken as a monopolar function with reflective symmetry in both the normal (radial at the midplane) and bi-normal (approximately poloidal) directions with the origin taken at the centre of the filament, where the density is initialized as a Gaussian function. The density and potential can be split into components with an even and odd parity about the filament centre in the bi-normal direction. These are labelled as $n^{+}$ and $\phi^{+}$ for the even components and $n^{-}$ and $\phi^{-}$ for the odd components. The density is assumed to take the form of an even function such that $n = n^{+}$ with $n^{-} = 0$. The electrostatic potential is a superposition of both parity components, $\phi = \phi^{+} + \phi^{-}$. This deconstruction is described schematically in figure \ref{Fig:Sym_schem}. 
Substituting the expressions for $n$ and $\phi$ in terms of $n^{+}$, $\phi^{+}$ and $\phi^{-}$, into equation \ref{Eqn:Vort} and noting that even order differential operators in the poloidal direction retain the parity of the function being operated on whilst odd order operators reverse the parity, gives
\begin{equation}
\label{Eqn:Vort_odd_even}
\rho_{s}^{2}\left[\frac{\partial\nabla_{\perp}^{2}\phi^{+}}{\partial t} + \textbf{v}_{E}^{-}\cdot\nabla\nabla_{\perp}^{2}\phi^{-} + \textbf{v}_{E}^{+}\cdot\nabla\nabla_{\perp}^{2}\phi^{+}\right] -\frac{\sigma_{||}T_{e}}{n}\left(\nabla_{||}^{2}\ln\left(n\right) - \nabla_{||}^{2}\phi^{+}\right) = 
\end{equation}
\[- \rho_{s}^{2}\left[\frac{\partial\nabla_{\perp}^{2}\phi^{-}}{\partial t} + \textbf{v}_{E}^{-}\cdot\nabla\nabla_{\perp}^{2}\phi^{+} + \textbf{v}_{E}^{+}\cdot\nabla\nabla_{\perp}^{2}\phi^{-}\right] + c_{s}\rho_{s}\bm{\xi}\cdot\nabla\ln\left(n\right) - \frac{\sigma_{||}T_{e}}{n}\nabla_{||}^{2}\phi^{-} 
\]
where the odd and even parity components of the $\textbf{E}\times\textbf{B}$ velocity are 
\begin{equation}
\textbf{v}_{E}^{-} = c_{s}\rho_{s}\textbf{b}\times\nabla\phi^{+} 
\end{equation}
\[\textbf{v}_{E}^{+} = c_{s}\rho_{s}\textbf{b}\times\nabla\phi^{-} \]
In equation \ref{Eqn:Vort_odd_even} all even parity terms have been placed on the LHS whilst all odd parity terms are on the RHS. For equation \ref{Eqn:Vort_odd_even} to be generally true both sides must be independently equal to $0$. This allows equation \ref{Eqn:Vort_odd_even} to be split into two coupled equations for the odd and even components of the potential. These are
\begin{equation}
\label{Eqn:Vort_odd} 
\rho_{s}^{2}\left[\frac{\partial\nabla_{\perp}^{2}\phi^{-}}{\partial t} + \textbf{v}_{E}^{-}\cdot\nabla\nabla_{\perp}^{2}\phi^{+} + \textbf{v}_{E}^{+}\cdot\nabla\nabla_{\perp}^{2}\phi^{-}\right] = c_{s}\rho_{s}\bm{\xi}\cdot\nabla\ln\left(n\right) - \frac{\sigma_{||}T_{e}}{n}\nabla_{||}^{2}\phi^{-} 
\end{equation}
for $\phi^{-}$ and 
\begin{equation}
\label{Eqn:Vort_even}
\rho_{s}^{2}\left[\frac{\partial\nabla_{\perp}^{2}\phi^{+}}{\partial t} + \textbf{v}_{E}^{-}\cdot\nabla\nabla_{\perp}^{2}\phi^{-} + \textbf{v}_{E}^{+}\cdot\nabla\nabla_{\perp}^{2}\phi^{+}\right] = \frac{\sigma_{||}T_{e}}{n}\left(\nabla_{||}^{2}\ln\left(n\right) - \nabla_{||}^{2}\phi^{+}\right)
\end{equation}
for $\phi^{+}$. The inclusion of the $\nabla_{||}n$ term in parallel Ohm's law provides a source for the even component of potential. It is commonplace in 2D theories of blob motion to neglect this term \cite{KrashenninikovPLA2001,D'IppolitoCPP2004,YuPoP2003,GarciaPoP2006,YuPoP2006,D'IppolitoReview}, though it is possible to accomodate it within 2D models \cite{BisaiPoP2005,MyraNF2013}. If the electrostatic potential is initialized as $0$ or an odd function, again commonplace in 2D simulations, then equation \ref{Eqn:Vort_even} prevents any growth of the even component in the absence of the $\nabla_{||}n$ term and the potential can only adopt an odd symmetry; in this sense simulations neglecting $\nabla_{||}n$ are a rather specific class of models which do not permit a breaking of the odd symmetry of the potential (without the addition of extra physics beyond the basic models). It should be noted that if the even component of the potential is of a similar magnitude to the odd component then significant motion of the filament can occur away from the radial direction. This violates the assumption that $n^{-} = 0$ and the present analysis breaks down. Never the less this serves to illustrate in a general manner the role played by 3D effects in the motion of filaments. By balancing the source for the odd component against the source for the even component a critical temperature can be approximated above which the initial growth of the even component is comparable to the odd component and 3D effects may impact the motion of the filament. Taking $\nabla_{||} \sim 1/L_{||}$ and $\nabla_{\perp}\sim 1/\delta_{\perp}$ as approximations where $L_{||}$ is the parallel length scale of the filament and $\delta_{\perp}$ is the perpendicular radius of the filament, the critical electron temperature is estimated as
\begin{equation}
\label{Eqn:Te_crit}
\frac{2c_{s}\rho_{s}\left|\bm{\xi}\right|\ln(n)}{\delta_{\perp}} \sim \frac{T_{e}\sigma_{||}}{nL_{||}^{2}}\ln(n) \Rightarrow T_{e} \approx \left(\frac{2nm_{e}L_{||}^{2}\left|\bm{\xi}\right|}{1.96\times 3\times 10^{10}\delta_{\perp}eB}\right)^{2/3} \approx 4eV
\end{equation}
where $\left|\bm{\xi}\right| = 1/R $ with the major radius $R = 1.5$m, $\delta_{\perp} = 2$cm, $n = 5\times 10^{18}$m$^{-3}$ and $B = 0.3$T are taken as representative values of the MAST SOL \cite{DudsonPPCF2008}, whilst $L_{||} = 8m$ is taken directly from the simulation geometry used here (see section \ref{Sec:Geom}). The numerical coefficients in the denominator of equation \ref{Eqn:Te_crit} arise from the definition of $\tau_{e}$ and $\eta_{||}$. This serves as an approximate temperature above which 3D effects can be expected to play a role in filament dynamics. Given that near SOL temperatures (and in some cases far SOL temperatures) in MAST are universally above $4$eV \cite{WalkdenRSI2015} it is important to assess the role of these 3D effects. To investigate the role of 3D effects on filament motion and the resulting particle transport scans in both electron temperature and filament radius have been conducted in a geometry representative of the SOL in MAST. 

\section{MAST Geometry}
\label{Sec:Geom}
\begin{figure}[htbp]
\centering
\includegraphics[width=\textwidth]{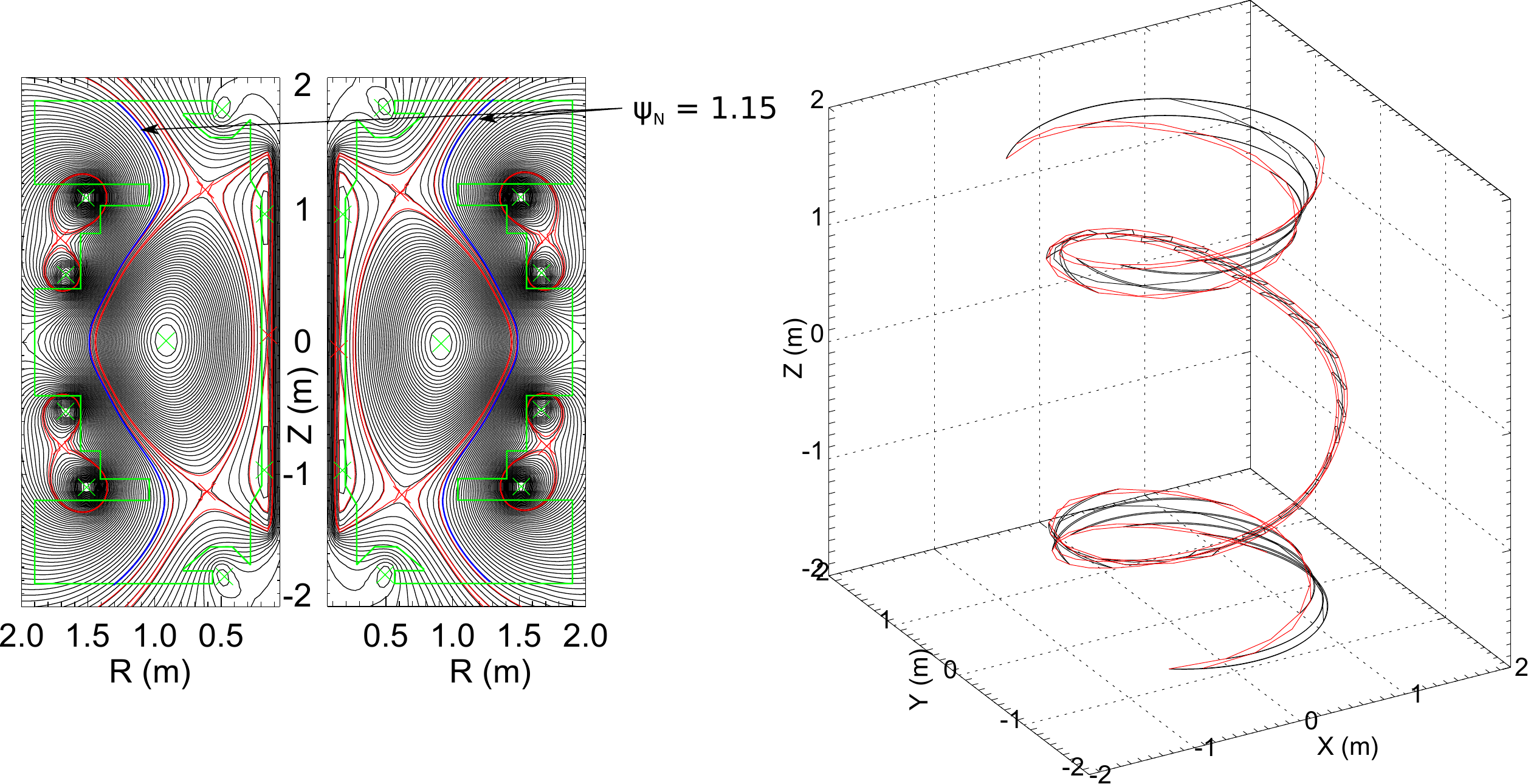}
\caption{Left: $\psi_{N} = 1.15$ flux surface in a MAST connected double-null Ohmic L-mode plasma. Right: 3D projection of the flux-tube domain constructed around a field line on the $\psi_{N} = 1.15$ surface}
\label{Fig:Flux_surface}
\end{figure}

The simulations in this paper have been conducted within the BOUT++ framework \cite{BOUT++Paper,BOUT++2014}. The simulation geometry is a flux-tube in the SOL of MAST constructed around a field line on the $\psi_{N} = 1.15$ flux surface where $\psi_{N} = \left(\psi - \psi_{sep}\right)/\left(\psi_{axis} - \psi_{sep}\right)$ is the normalized poloidal magnetic flux and $\psi$, $\psi_{axis}$ and $\psi_{sep}$ are the poloidal magnetic flux on the flux surface, magnetic axis and separatrix respectively. $\psi_{N} = 1.15$ gives a flux surface suitably far from the separatrix to consider filaments as isolated structures, but sufficiently far from the first wall to study their motion. Figure \ref{Fig:Flux_surface} shows the $\psi_{N} = 1.15$ flux surface in MAST, as well as a 3D projection of the flux-tube domain. 
 Limitations of the flux-tube approach used here are discussed in \cite{WalkdenThesis}.
\\The simulations are carried out in the field-aligned ballooning coordinate system defined by the coordinates
\begin{equation}
\begin{array}{c c c c c}
x = \psi - \psi_{0} & & y = \theta & & z = \zeta - \int_{\theta_{0}}^{\theta}\nu\left(\psi,\theta'\right) d\theta'
\end{array}
\end{equation}
Fourth order central differencing is applied to diffusive operators in the $x$ and $y$ dimensions whilst a 3rd order WENO (Weighted Essentially Non-Oscillatory) scheme is used for parallel advective operators \cite{JiangWENOPaper}. FFTs are used in the angular coordinate $z$. Neumann boundary conditions are applied to all variables on the boundaries in $x$ whilst periodic boundary conditions are applied in $z$ with a toroidal period of $15$. This allows filaments to be simulated with a toroidal mode number of 15, which is on the lower bound of the experimentally observed filament quasi-periodicity \cite{DudsonPPCF2008}. This periodicity was chosen as it allowed filaments to be treated as isolated; their profiles in density or potential did not reach the periodic boundary, particularly as the potential dipole was established. The $y$ coordinate labels the position along the magnetic field line. Sheath boundary conditions are applied to the boundaries in $y$ by setting the parallel current $J_{||}$ to the current at the entrance to the magnetic pre-sheath $J_{sh}$ \cite{NedospasovNF85,LoizuPoP2012} where
\begin{equation} 
J_{sh}^{\pm} = \pm nec_{s}\left(1-\exp\left[\phi_{W}-\phi^{\pm}\right]\right)
\end{equation}
with the wall potential, $\phi_{W}$, set to $0$ since it acts as a constant reference in the isothermal case. See \cite{BOUT++Paper,WalkdenPPCF2013} for more details concerning the coordinate system. The system is integrating in time using the PVODE \cite{PVODEPaper} implicit time-solver which is a Jacobian-free Newton-Krylov solver using the GMRES Krylov subspace method. $512$ grid points have been used in $x$, $256$ are used in $z$ and $64$ are used in the parallel $y$ coordinate. This provides sufficient resolution to capture the most unstable resistive drift-wave when dissipation is removed from the system \cite{WalkdenThesis}. Filaments are initialized with a 2D Gaussian density in the perpendicular plane with a peak density of twice the background. They are initially homogenous along the magnetic field line. The potential is initialized to zero and grows in the initial phase of the simulation. Unlike ref \cite{EasyPoP2014} the background is flat.
\\ \\As illustrated in ref. \cite{WalkdenPPCF2013} the magnetic geometry of the SOL gives rise intrinsically to gradients in the filament density along the magnetic field. There are two primary causes of this: 1. Ballooning of the filament near the midplane \cite{WalkdenPPCF2013} and 2. Enhanced dissipation in the divertor region \cite{WalkdenThesis}. Both of these effects are a result of the variation in the magnetic geometry of the flux-tube  along the field line. In the divertor the shearing and contraction of the flux tube in the bi-normal direction lead to enhanced cross-field density gradients. As a result the diffusion of density is strongly enhanced in this region, even at dissipation levels well below that which may be expected experimentally \cite{FundamenskiNF2007}. By providing mechanisms for the formation of parallel density gradients the magnetic geometry naturally drives the source for the even component of the potential.

\section{Dynamics of the filament cross-section}
\label{Sec:Dyn}
\subsection{Effect of the electron temperature}
\begin{figure}[htbp]
\centering
\includegraphics[width=\textwidth]{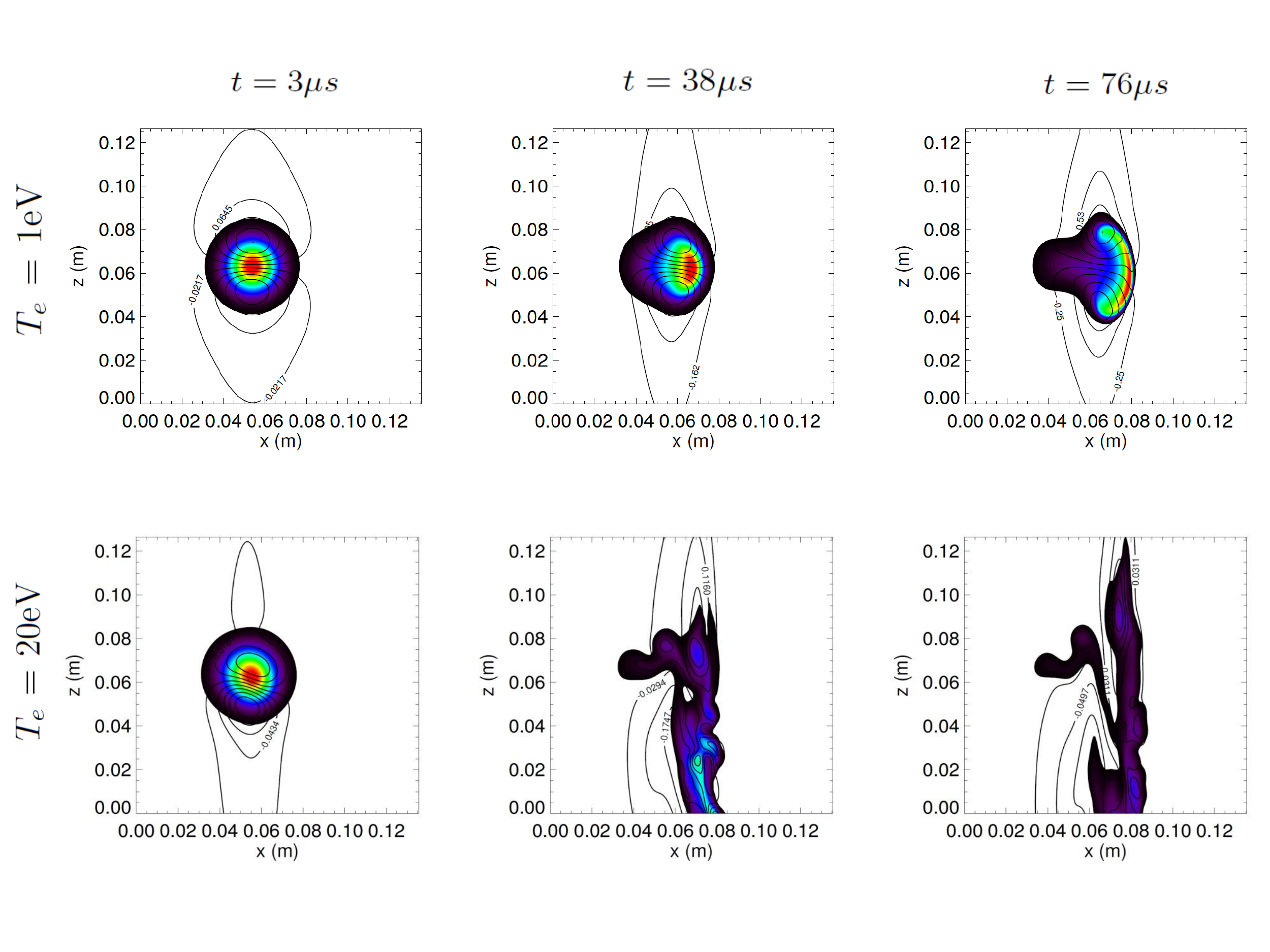}
\caption{Top: Filament cross-section in the perpendicular plane at the outboard midplane with $T_{e} = 1$eV. Bottom: As top but with $T_{e} = 20$eV. Colour represents density and contours represent electrostatic potential. Increasing the electron temperature increases the prominence of the Boltzmann source which leads to a departure from interchange dynamics.}
\label{Fig:Tem_Xsecs}
\end{figure}

\begin{figure}[htpb]
\centering
\includegraphics[width=0.8\textwidth]{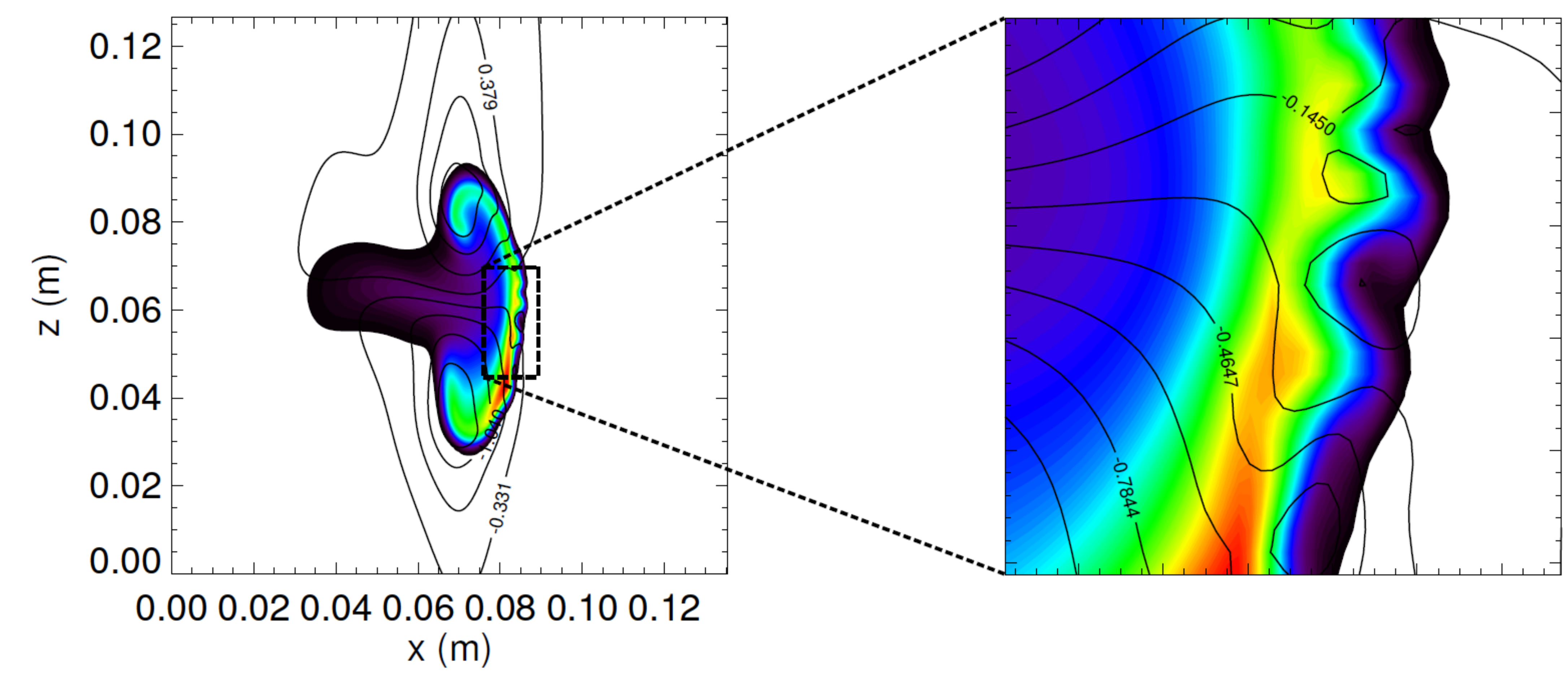}
\caption{Filament cross-section at the outboard midplane with $T_{e} = 1$eV 113$\mu$s after initialization. Resistive drift-waves form on the front of the filament due to density gradient build up, but do not significantly affect the filament. Their length scale is much smaller than that of the filament.}
\label{Fig:drift_waves}
\end{figure}

\begin{figure}[htpb]
\centering
\includegraphics[width=0.8\textwidth]{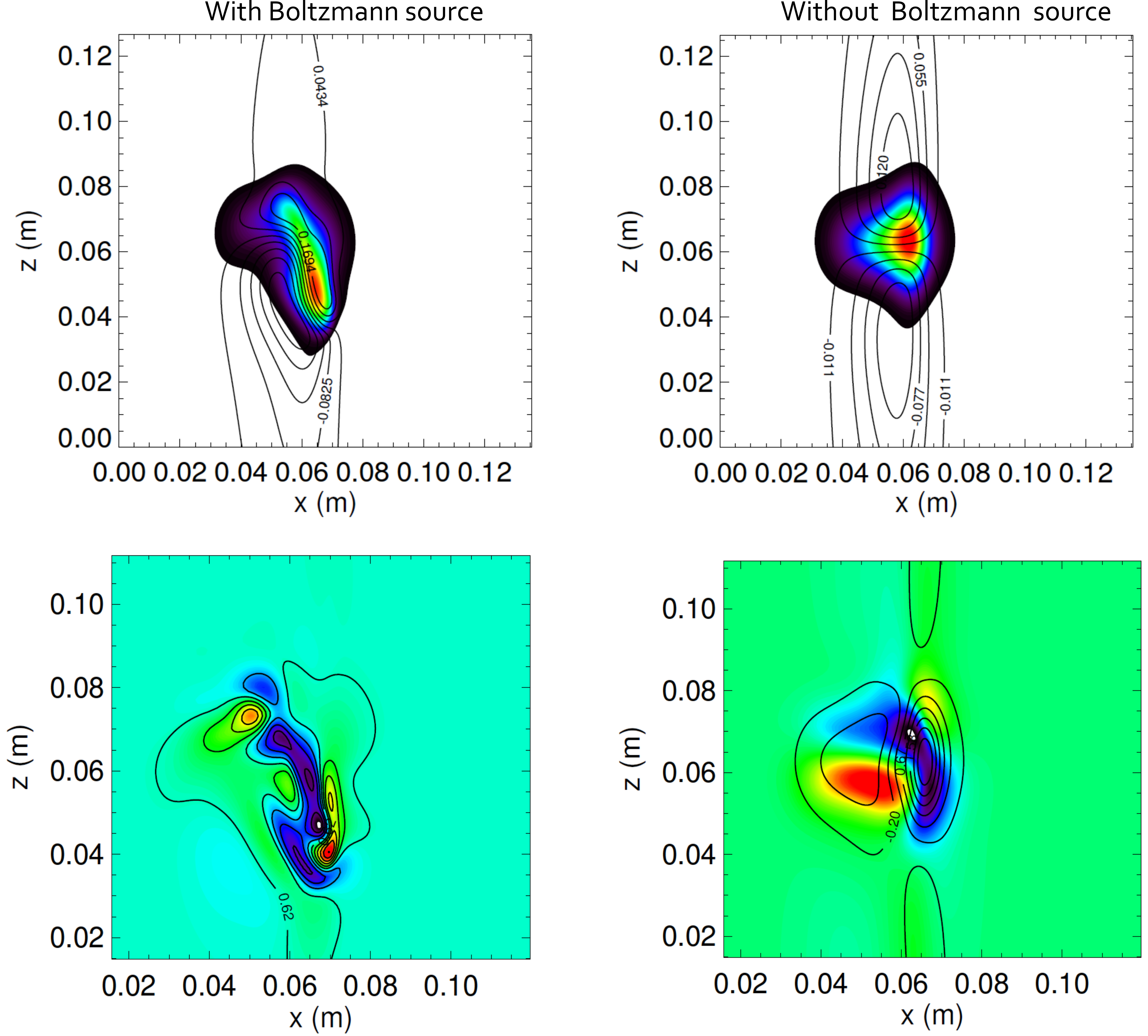}
\caption{Top: Filament cross-section at the outboard midplane with (left) and without (right) the Boltzmann source included. Bottom: $\nabla_{||}\ln(n)$ (colour) and $\nabla_{||}\phi$ (contours) for the cases with and without the Boltzmann source included. The Boltzmann source drives the filament towards a Boltzmann response and is responsible for the observed dipole rotation.}
\label{Fig:No_Boltz}
\end{figure}

\begin{figure}[htbp]
\centering
\includegraphics[width=0.8\textwidth]{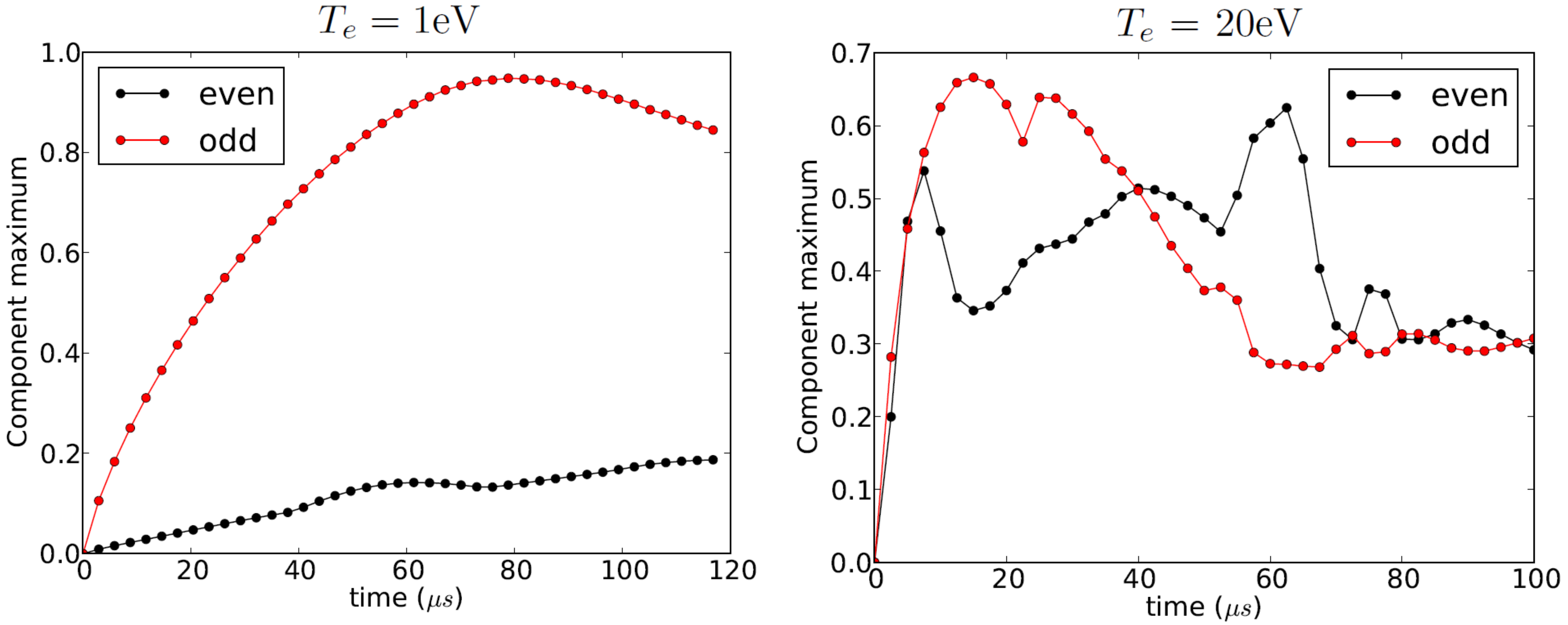}
\caption{Absolute maximum of the even and odd components of potential (normalized to the absolute maximum of the total potential over the entire simulation) during the evolution of the $1$eV and the $20$eV filaments. In the $1$eV case the growth of the odd component is significantly faster than the even. In the $20$eV case the two are comparable.}
\label{Fig:Sym_comps_Te}
\end{figure}

\begin{figure}[htbp]
\centering
\includegraphics[width=0.8\textwidth]{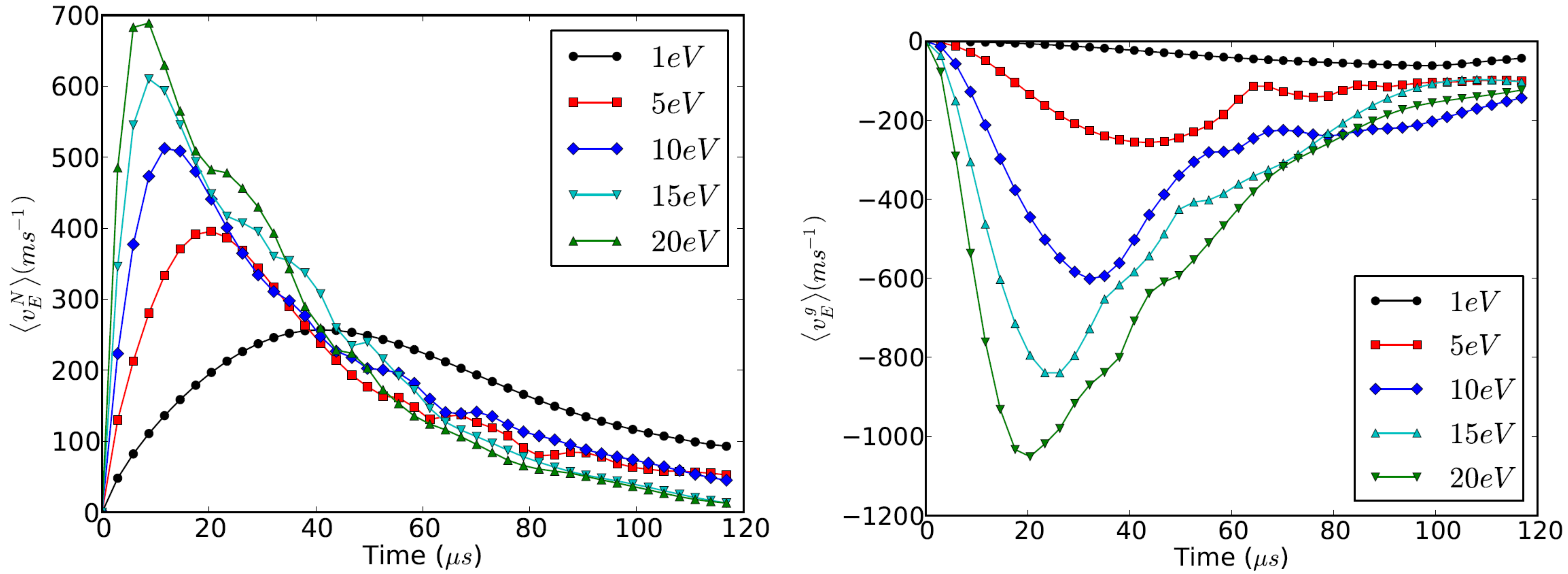}
\caption{Normal (left) and bi-normal (right) centre of mass velocity of the filament cross-section at the outboard midplane over the scan in $T_{e}$.}
\label{Fig:v_Tescan}
\end{figure}

\begin{figure}[htbp]
\centering
\includegraphics[width=\textwidth]{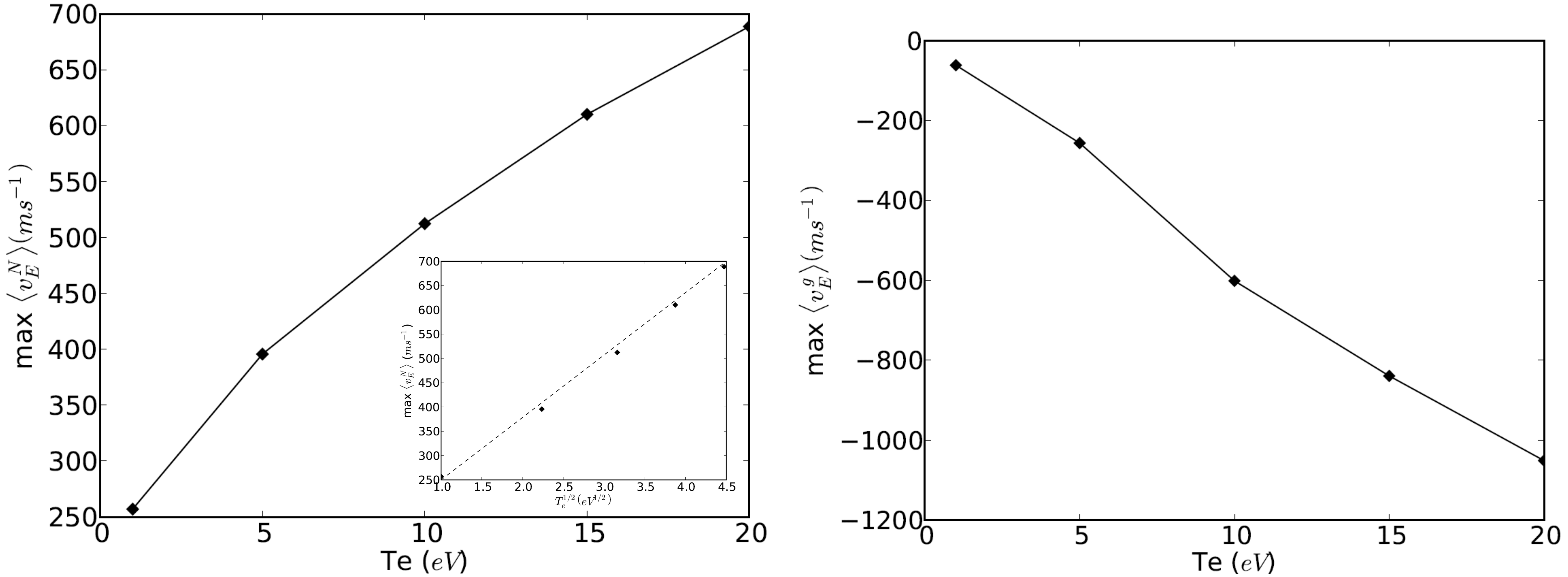}
\caption{Maximal velocity in the normal (left) and bi-normal (right) direction. Inset is the maximal normal velocity as a function of $T_{e}^{1/2}$ demonstrating a reasonably linear relationship.}
\label{Fig:vmax_Tescan}
\end{figure}

\begin{figure}[htbp]
\centering
\includegraphics[width=0.8\textwidth]{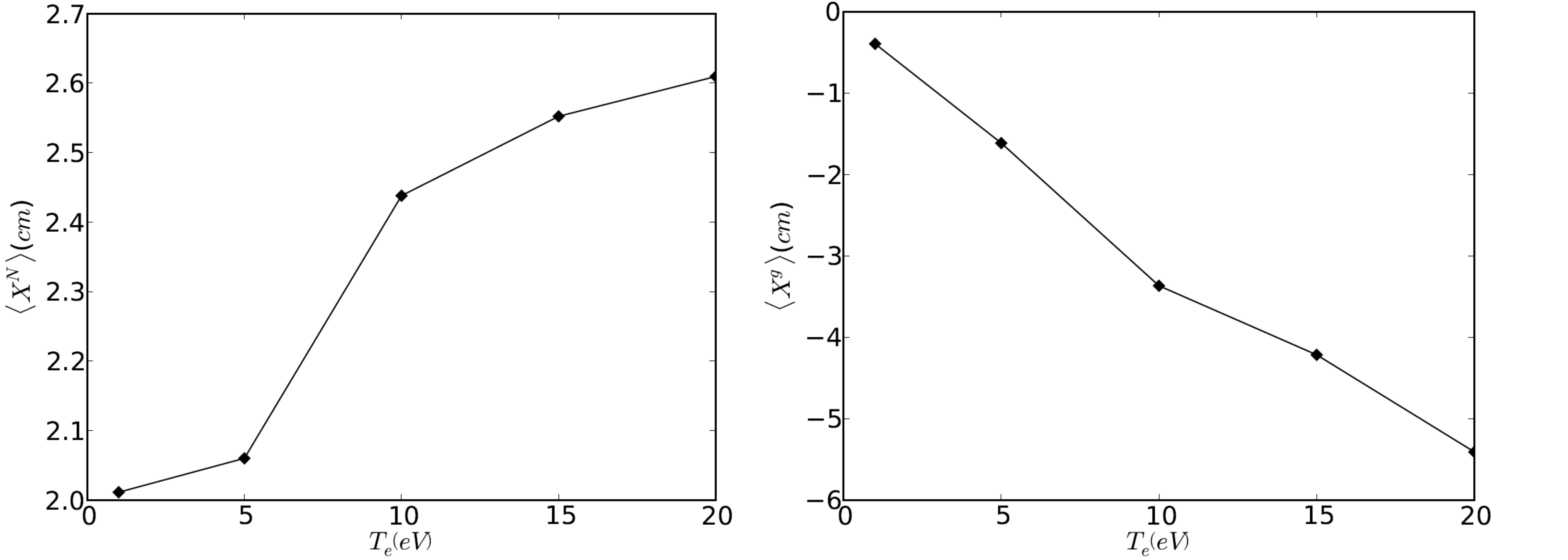}
\caption{Net displacement of the filament cross-section at the outboard midplane in the normal (left) and bi-normal (right) direction calculated by integrating the velocity curves in figure \ref{Fig:v_Tescan} over the simulation time.}
\label{Fig:X_Tescan}
\end{figure}

Figure \ref{Fig:Tem_Xsecs} shows the evolution of the filament cross-section at the midplane with constant background temperatures $T_{e} = 1$eV and $T_{e} = 20$eV. 
The dynamics of the filament vary dramatically between the two cases. At $1$eV the behaviour of the filament is reminiscent of the inertially limited regime found in 2D models \cite{MyraPoP2006} where a dominantly dipolar potential causes the filament to 'mushroom' outwards. Resistive drift-waves are observed to form on the filament front during the latter part of the filament's motion at 1eV, as described by Angus \emph{et.al} \cite{AngusPRL2012,AngusPoP2012}. These instabilities, shown in figure \ref{Fig:drift_waves}, form when density gradients on the filament front steepen as a result of its motion, and have a typical length scale much smaller than that of the filament cross-section. 
\\ \\At $20$eV the dynamics of the filament depart from the $1$eV case. The dipole potential is not sustained, instead showing an initial rotation before a break down towards smaller length scales. The rotation of the dipole leads to motion in the poloidal direction,  similar to that observed in refs. \cite{AngusPoP2012,MyraNF2013}, before the break down of the potential acts to disperse the filament. To show that these features are a result of the inclusion of the $\nabla_{||}n$ term in equation \ref{Eqn:Vort_even} the $20$eV simulation has been re-run with this term now excluded. Figure \ref{Fig:No_Boltz} compares the filament after $12\mu$s of evolution with and without the $\nabla_{||}n$ term.
Removing the $\nabla_{||}n$ term allows the filament to return to the dipole potential by removing any source for an even component of the potential. 
\\ \\Increasing temperature decreases the resistivity of the plasma by $T_{e}^{-3/2}$ (thus a $20$ times increase in $T_{e}$ corresponds to a $90$ times decrease in resistivity). Equation \ref{Eqn:Jpar} suggests that as resistivity is decreased the filament should adopt a Boltzmann response such that $\nabla_{||}\ln(n) \sim \nabla_{||}\phi$. This has been tested in the bottom frames of figure \ref{Fig:No_Boltz} and, as expected, the case where $\nabla_{||}n$ is included in the system does indeed tend towards a Boltzmann response, whilst the counter case does not. Notably this shows that 2D models which inherently neglect the $\nabla_{||}n$ term are not compatible with the requirement for the plasma to obtain a Boltzmann response, even at relatively modest temperatures such as $20$eV. As a result care should be taken when applying 2D models to situations where even modest parallel density gradients can occur. 
\\ \\The even and odd parity components of the potential, $\phi^{+}$ and $\phi^{-}$ can be extracted from the full potential solution, $\phi$, using
\begin{equation}
\phi^{+} = \frac{1}{2}\left(\phi\left(z - z_{0}\right) + \phi\left(z_{0}-z\right)\right)
\end{equation}
\begin{equation}
\phi^{-} = \frac{1}{2}\left(\phi\left(z-z_{0}\right) - \phi\left(z_{0}-z\right)\right)
\end{equation}
 where $z$ is the bi-normal coordinate and $z_{0}$ is the centre of the filament in the perpendicular plane.
Figure \ref{Fig:Sym_comps_Te} shows the evolution of the absolute maximum of the odd and even components of the potential during the simulations of the $1$eV and $20$eV filaments. 
\\ \\The relative growth of the even component of the potential increases dramatically with increasing electron temperature due to the increased strength of the Boltzmann source. In the $20$eV case the poloidal motion breaks the even symmetry of the density early in the simulation, which causes the arguments based on parity to break down. The initial growth of the two components of potential occurs before this symmetry breaking though and confirms that increasing electron temperature increases the drive for the even component of potential. 
\\ \\A useful quantity for assessment of the filament motion is the velocity in the normal ($\nabla\psi/\left|\nabla\psi\right|$) and bi-normal ($\textbf{b}\times\nabla\psi / \left|\textbf{b}\times\nabla\psi\right|$) directions respectively. These are calculated by projecting the $\textbf{E}\times\textbf{B}$ velocity onto the normal and bi-normal unit vectors to give
\begin{equation}
v_{E}^{N} = \textbf{v}_{E}\cdot\hat{\textbf{e}}_{N} = \textbf{v}_{E}\cdot \frac{\nabla\psi}{RB_{\theta}}
\end{equation}
and
\begin{equation}
v_{E}^{g} = \textbf{v}_{E}\cdot\hat{\textbf{e}}_{g} = \textbf{v}_{E}\cdot\left(\nabla\psi\times\textbf{b}\right) = \textbf{v}_{E}\cdot\frac{RB_{\theta}}{B}\left(\nabla z + I\nabla\psi\right)
\end{equation}
where $v_{E}^{N}$ is the velocity in the normal direction and $v_{E}^{g}$ is the velocity in the bi-normal direction, $R, B_{\theta}$ and $I$ are the major radius, poloidal magnetic field strength and integrated magnetic shear of the flux tube. The convention adopted in ref \cite{RyutovPoP2006} where the bi-normal direction is labelled with $g$ is adopted here. The centre of mass velocity of the filament can then be calculated from 
\begin{equation}
\langle v_{E}^{N,g}\rangle = \int\int v_{E}^{N,g}\delta n dx_{N}dx_{g} / \left(\int\int \delta n dx_{N}dx_{g}\right)
\end{equation}
where $\delta n$ is the density of the filament on top of the background.  Figure \ref{Fig:v_Tescan} shows the evolution of $\langle v_{E}^{N}\rangle$ and $\langle v_{E}^{g}\rangle$ for each filament in the temperature scan  with $\delta n/n_{0} \approx 2$ and $\delta_{\perp} = 4cm$.
The velocity in each case shows an initial acceleration period as the potential grows, followed by a peak and a deceleration period. In both the normal and bi-normal direction the maximum speed attained by the filament increases with electron temperature. In figure \ref{Fig:vmax_Tescan} the maximum speed is given as a function of electron temperature.
Figure \ref{Fig:vmax_Tescan} indicates that $\langle v_{E}^{N}\rangle \propto T_{e}^{1/2}$ whilst $\langle\left|v_{E}^{g}\right| \rangle \propto T_{e}$. The scaling of the normal velocity is consistent with the  inertial regime of filament motion \cite{OmotaniArXiV2015} where the characteristic velocity has the scaling
\begin{equation}
v^{N} \sim c_{s}\sqrt{2\left|\bm{\xi}\right|\delta_{\perp}} \propto T_{e}^{1/2}\delta_{\perp}^{1/2}
\end{equation}
A scaling for the bi-normal velocity can be derived by balancing order of magnitude terms in equation \ref{Eqn:Vort_even}. If the vorticity term, which acts to mix even and odd components of the potential, is balanced against the drive term and the odd component of the potential is taken to be inertially limited (as suggested by the present simulations) the binormal velocity obtains the scaling
\begin{equation}
v_{g} \sim \frac{c_{s}\sigma_{||}B}{\delta n e L_{||}}\sqrt{\frac{\delta_{\perp}}{8\left|\bm{\xi}\right|}} \propto T_{e}^{2}\delta_{\perp}^{1/2}
\end{equation}
On the other hand if the drive term is ballanced by the parallel dissipation term such that the Boltzmann response is satisfied the bi-normal velocity obtains the scaling
\begin{equation}
v_{g} \sim \frac{c_{s}\rho_{s}}{\delta_{\perp}}\ln\left(\delta n\right) \propto T_{e}\delta_{\perp}^{-1}
\end{equation}
The trend of the maximal bi-normal velocity observed over the temperature scan, shown in figure \ref{Fig:vmax_Tescan}, indicates that the Boltzmann response scaling is appropriate to describe the filaments observed here.
\\ \\Despite the increase in the peak velocity, a decrease in the time taken to reach the velocity peak and a speed-up in the deceleration following the peak is observed. These factors all affect the net displacement of the filament and may cause it to scale differently to the velocity. Figure \ref{Fig:X_Tescan} shows the net displacement of the filament centre of mass in the normal and bi-normal directions.
\\ \\The linear increase in the bi-normal velocity with temperature translates into a linear increase in the bi-normal displacement of the filament by $\approx 5$cm across the temperature scan. By contrast the increase in the normal displacement is weaker than the corresponding increase in velocity. This is because, although the peak velocity of the filament increases with temperature, the deceleration after the peak is also quicker as the break down into drift-wave turbulence occurs. As a result a much more modest $\approx 0.6$cm increase in the normal displacement occurs across the temperature scan.

\subsection{Filament Width Scan}
\begin{figure}[htbp]
\centering
\includegraphics[width=1.1\textwidth]{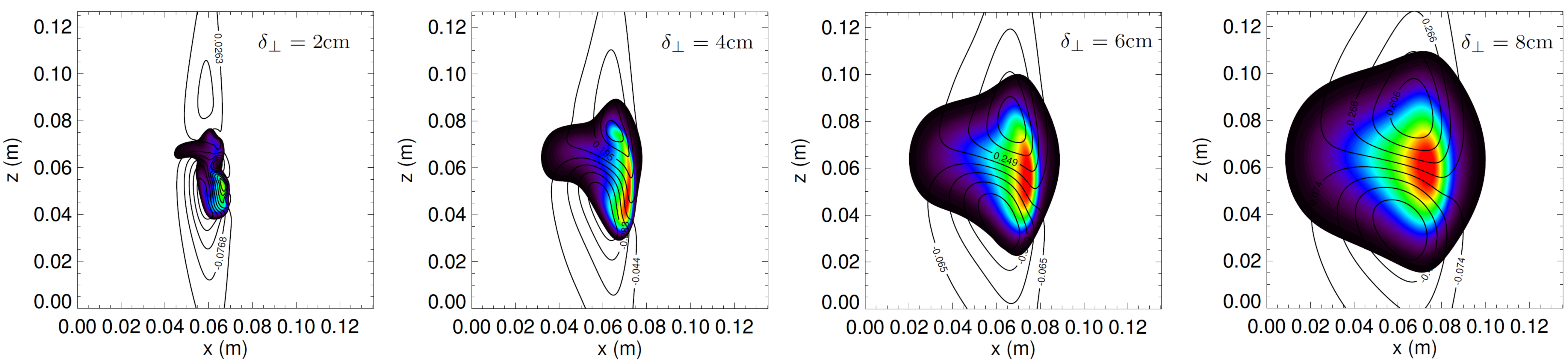}
\caption{Filament cross-sections sampled 38$\mu$s after initialization for varying values of the initial cross-sectional width (in the bi-normal direction) $\delta_{\perp}$. The prevelance of 3D effects reduces as the filament width increases from left to right.}
\label{Fig:dscan_Xsecs}
\end{figure}

\begin{figure}[htbp]
\centering
\includegraphics[width=0.8\textwidth]{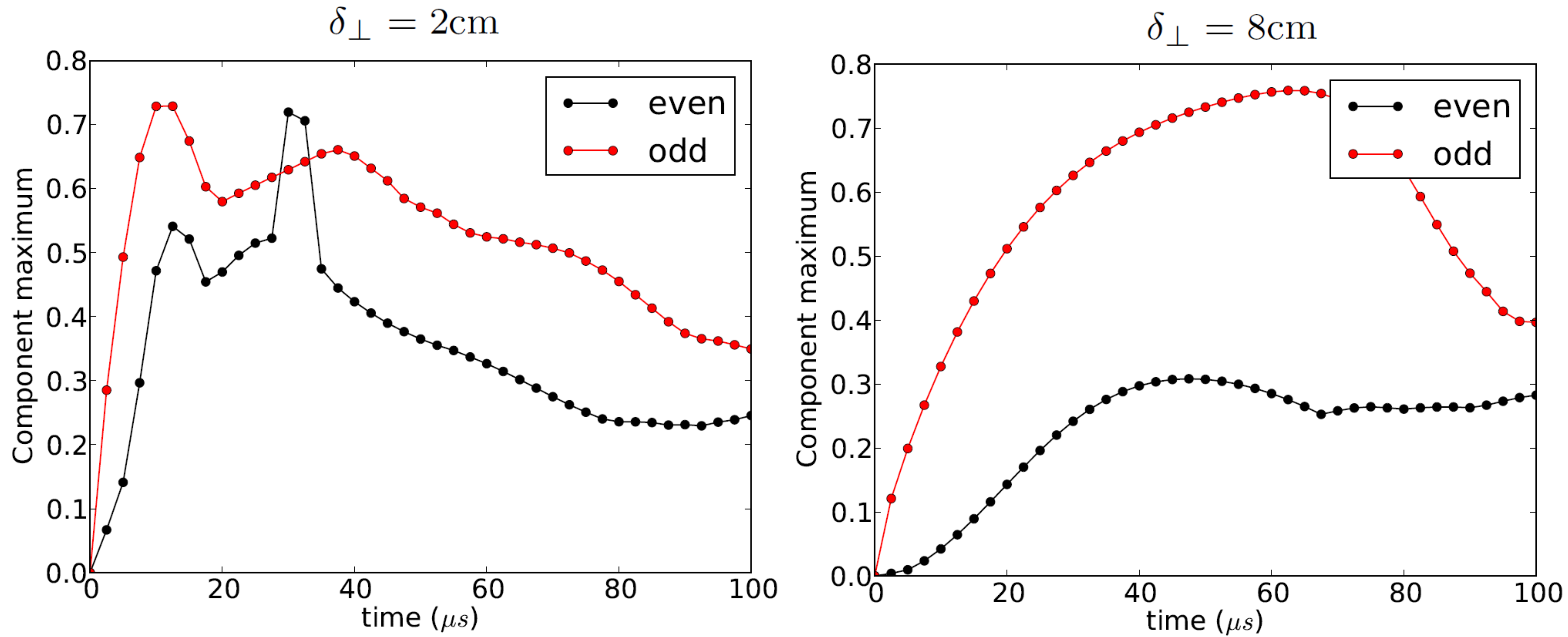}
\caption{Growth of the absolute maximum of the even and odd components of the potential at $\delta_{\perp} = 2$cm and $\delta_{\perp} = 8$cm. The growth of the two components is comparable in the smaller filament but the growth of the odd component is faster in the larger filament.}
\label{Fig:sym_comps_width}
\end{figure}

\begin{figure}[htbp]
\centering
\includegraphics[width=0.9\textwidth]{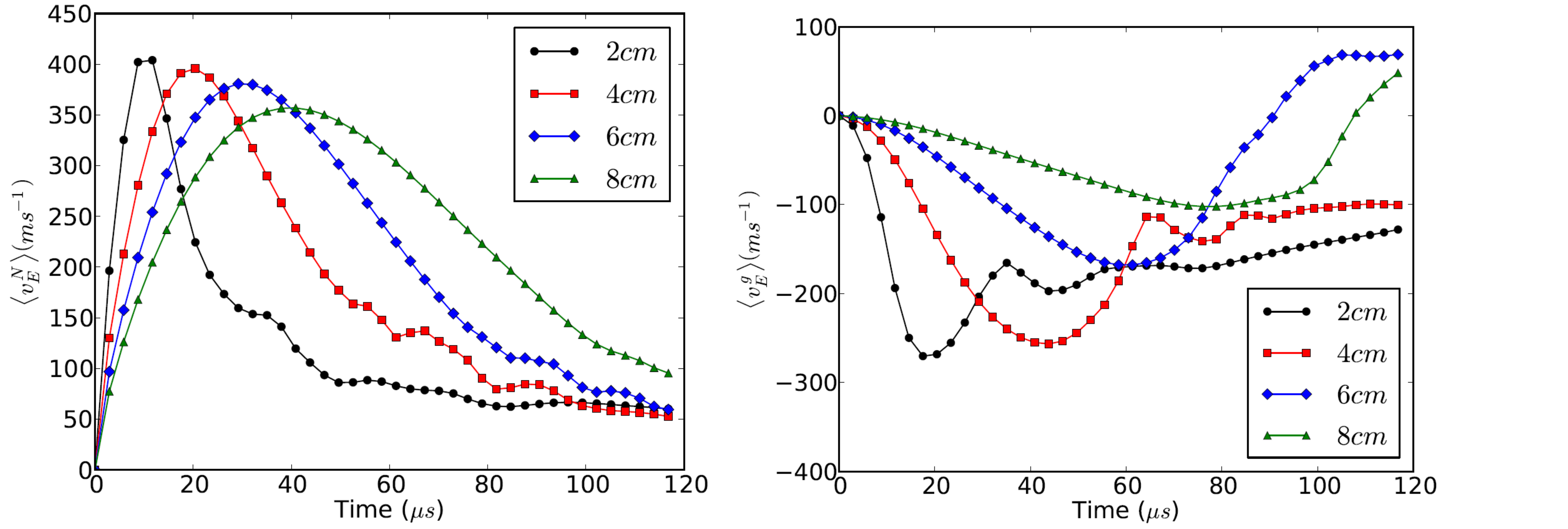}
\caption{Centre of mass velocity in the normal (left) and bi-normal (right) directions over the scan in $\delta_{\perp}$.}
\label{Fig:v_dscan}
\end{figure}

\begin{figure}[htbp]
\centering
\includegraphics[width=\textwidth]{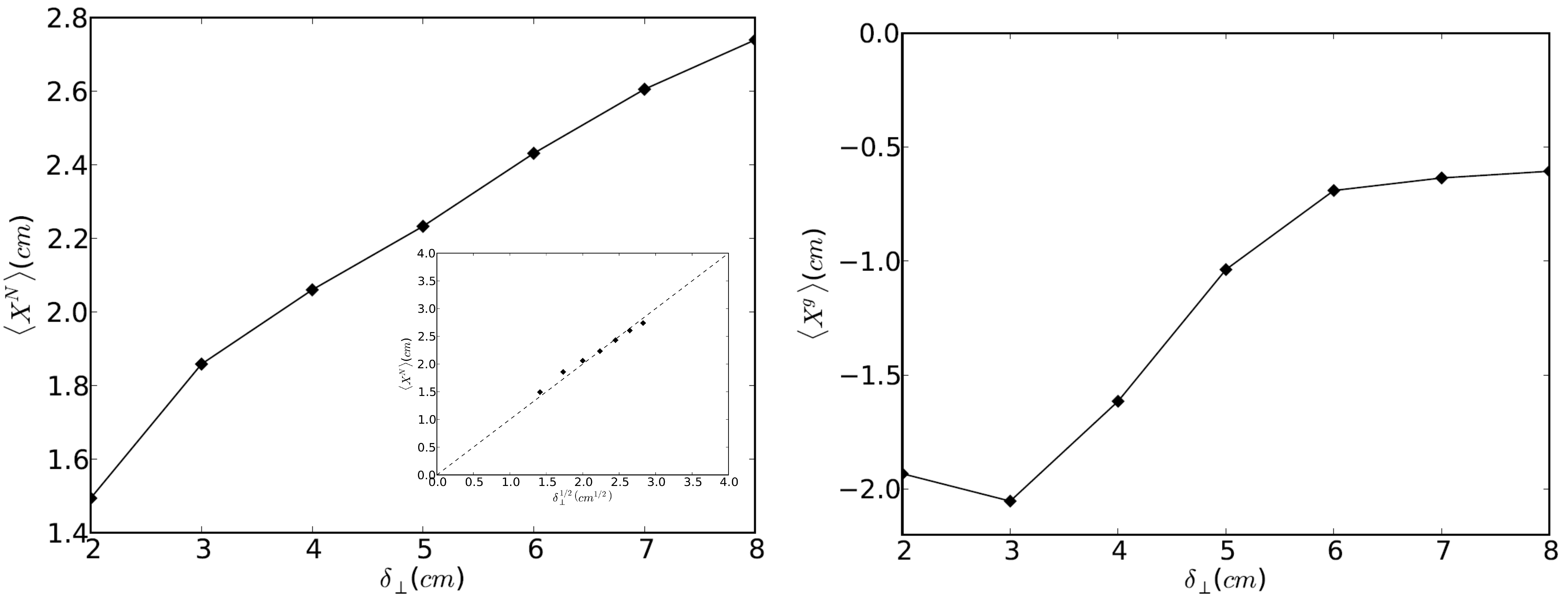}
\caption{Net displacement of the filament in the normal (left) and bi-normal (right) direction across the $\delta_{\perp}$ scan measured by integrating the curves shown in figure \ref{Fig:v_dscan} over the simulation time. Inset is the net displacement in the normal direction as a function of $\delta_{\perp}^{1/2}$ displaying an approximately linear relationship.}
\label{Fig:X_dscan}
\end{figure}

To compliment the scan in electron temperature and further investigate the dynamics of filaments in MAST a scan in filament width has been conducted in the range $2$cm to $8$cm at a constant electron temperature of $T_{e} = 5$eV , which is characteristic of the far SOL of MAST \cite{ElmorePPCF2012,WalkdenRSI2015},  and with $\delta n/n_{0} \approx 2$ held constant. This encompasses the range of filament widths found experimentally on MAST \cite{DudsonPPCF2008,AyedPPCF2009}. The quoted filament widths here refer to the width in the bi-normal direction at initialisation where the density exceeds a 1\% increase over the background (this is also the limit of the colour levels shown in any cross-section image of the filament). Figure \ref{Fig:dscan_Xsecs} shows the midplane cross-section of filaments in the $\delta_{\perp}$ scan. 
\\ \\The impact of 3D effects is stronger in smaller filaments, where the density shows strong bi-normal motion and the potential begins to break down into smaller scale structures. By contrast the larger filaments show a dominantly dipolar potential which leads predominantly radial motion. This suggests that,  for the parameters used here, larger filaments move mainly in the radial direction, whilst smaller filaments can have a significant bi-normal component to their motion. Figure \ref{Fig:sym_comps_width} shows the evolution of the even and odd components of the potential at the limits of the $\delta_{\perp}$ scan. 
\\ \\The relative growth of the even component of the potential decreases as the filament width increases. This occurs because at larger filament widths dissipation due to inertia diminishes quicker than the drive due to the diamagnetic current reduces. As a result there is comparatively more current flowing across the filament in larger filaments which must be closed by current flow along the magnetic field line, thereby increasing $J_{||}$. Since the resistivity, $\eta$, is constant Ohm's law (equation \ref{Eqn:Jpar}) implies that an increase in the parallel current will drive the filament away from the Boltzmann response. This then leads to the observed decrease in the relative strength of $\phi^{+}$ with respect to $\phi^{-}$ as $\delta_{\perp}$ is increased.
\\ \\Figure \ref{Fig:v_dscan} shows the centre of mass velocity of the filament at the midplane in the normal and bi-normal directions in a subset of simulations contained in the $\delta_{\perp}$ scan.
\\ \\The maximum velocity of the filament decreases weakly with increasing $\delta_{\perp}$ in the normal direction, in contrast to the $\delta_{\perp}^{1/2}$ scaling predicted in the inertially limited regime. The bi-normal velocity also shows significantly more complex behaviour than the $T_{e}$ scan. Whilst the peak velocity does decrease with increasing $\delta_{\perp}$, as predicted by the Boltzmann response scaling, it does not show a clear $\delta_{\perp}^{-1}$ dependence. This is likely due to the fact that the mechanism for bi-normal motion is more complex than for the normal motion and involves the break down of the potential into small scale structures. As a result this may not be particularly well represented by a characteristic velocity based on an order of magnitude scaling of a ballistic velocity. In fact the even component of potential cannot, on its own, lead to non-circulatory motion and requires some non-linear interaction with the odd component to produce bi-normal motion. Such complex motion is unlikely to adhere to simple scaling arguments.
\\ \\The net displacement of the filament is presented in figure \ref{Fig:X_dscan}. 
The displacement in the normal direction obtains a scaling of $\delta_{\perp}^{1/2}$ suggesting that the inertially limited regime may be well represented by the average velocity of the filament over its lifetime (i.e. displacement / simulation time). In the bi-normal direction the displacement, as with the velocity, shows a general decrease with increasing $\delta_{\perp}$ as predicted, however the detailed scaling is not clear. What is clear is that the radial propagation of the filament centre-of-mass increases with filament width and can outweigh the bi-normal displacement.  
\subsection{Comparison with the two-region model}
\begin{figure}[htbp]
\centering
\includegraphics[width=0.8\textwidth]{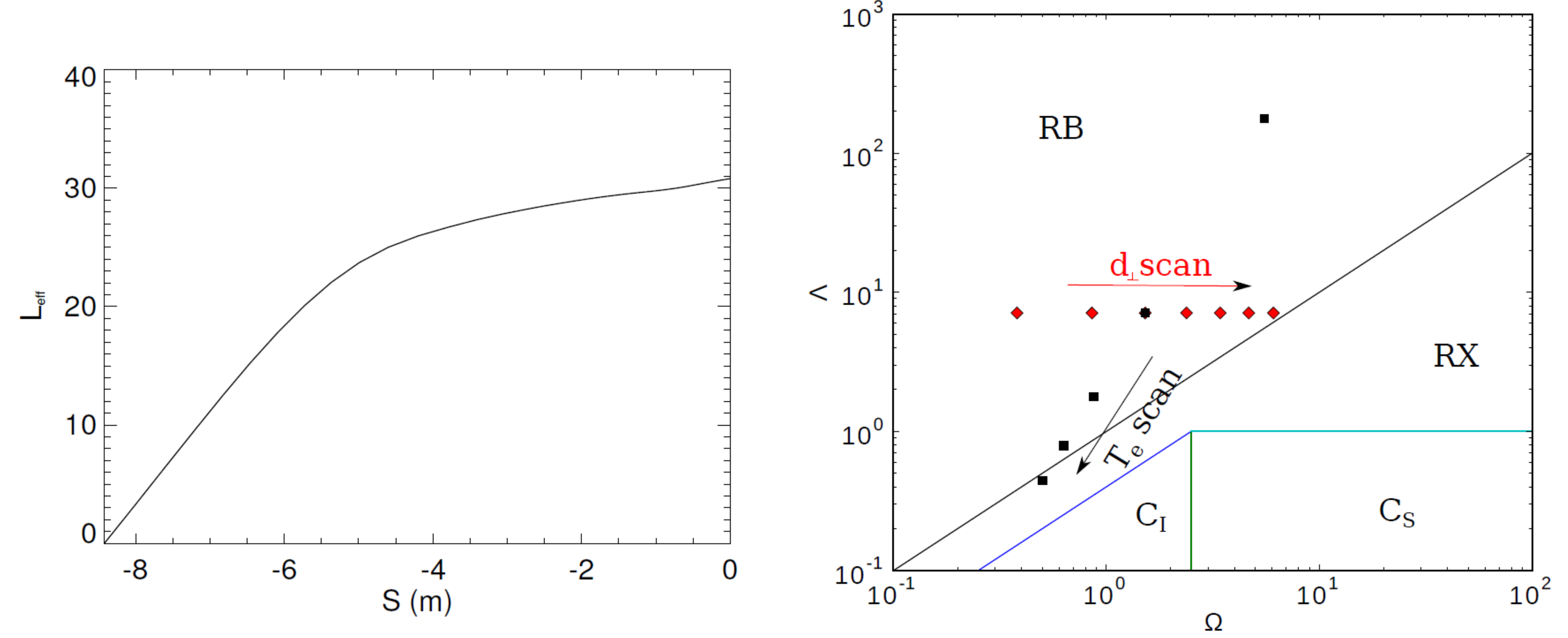}
\caption{Left: $L_{eff}$ calculated along the MAST SOL flux tube. Integration is carried out starting at the divertor. In the divertor region ($S < -5$m) $L_{eff}$ begins to depart dramatically from $L_{||}$. Right: Corrected paths in $\Lambda,\Omega$ space.}
\label{Fig:2region_2}
\end{figure}

The two-region model, developed by Myra and Russell \cite{MyraPoP2006, RussellPoP2007}, is an electrostatic 2D fluid model which includes both plasma resistivity and the effects of the magnetic geometry through a transformation of the cross-field differential operators. It has recently been extended to include the effects of warm ions \cite{ManzPoP2013}. By including geometrical effects the two-region model is relevant for comparison with the simulations carried out here. The (cold ion) two region model has four regimes of filament dynamics which are characterized by the dimensionless parameters $\Lambda$ and $\Omega$ where 
\begin{equation}
\label{Eqn:Lambda}
\Lambda = \frac{L_{||}\omega_{e}}{c_{s}\tau_{ei}\Omega_{i}}
\end{equation}
is a dimensionless resistivity and 
\begin{equation}
\label{Eqn:Omega}
\Omega = \left(\frac{\delta_{\perp}}{\delta^{*}}\right)^{5/2}
\end{equation}
which represents the blob width. Each regime is characterized by a different closure/dissipation mechanism for the diamagnetic (driving) current across the filament. The four regimes of dynamics, their current closure mechanism and their velocity scalings with $T_{e}$ and $\delta_{\perp}$ are given in table \ref{Tbl:two_region}. For a derivation of the model and details see ref \cite{MyraPoP2006}.
\begin{table}[htbp]
\centering
\begin{tabular}{ c c c }
\hline \hline
Regime & Closure Mechanism & Velocity Scaling \\ \hline \hline \\
Resistive Ballooning (RB) & Local polarization currents & $v 
\propto T_{e}^{1/2}\delta_{\perp}^{1/2} $\\
Resistive X-point (RX)& X-point polarization currents &  $v 
\propto T_{e}^{-1/2}\delta_{\perp}^{-2}$ \\
Sheath Interchange (C$_{s}$)& Sheath currents & $v 
\propto T_{e}^{3/2}\delta_{\perp}^{-2}$ \\
Ideal Interchange (C$_{I}$)& Global polarization currents & $v 
\propto T_{e}^{1/2}\delta_{\perp}^{1/2} $ \\ \\\hline \hline
\end{tabular}
\caption{Scaling of the normal velocity in the two-region model under differing regimes of current dissipation.}
\label{Tbl:two_region}
\end{table}
The resistive ballooning regime mentioned above is identical the the regime termed the inertially limited regime here. Both the $T_{e}$ and $\delta_{\perp}$ scans trace specific paths in $\Lambda$, $\Omega$ space. In calculating these paths the filament width has been converted from the definition used here (1\% above background) to the Gaussian width of the filament, which is more appropriate for comparison with the two-region model. Geometric parameters are calculated directly from the flux-tube geometry. The least well constrained parameter required for this comparison is the parallel connection length, $L_{||}$. In the two-region model $L_{||}$ occurs when an integration of the curvature drive is carried out along the magnetic field line and balanced against either parallel or sheath currents. The curvature drive is assumed to be fixed at the midplane value such that 
\begin{equation}
\int_{0}^{L_{||}} 2c_{s}\rho_{s}\bm{\xi}\cdot\nabla \ln\left(n\right) dL \approx \frac{2c_{s}\rho_{s}\left|\bm{\xi}\right|_{mid}}{\delta_{\perp}} L_{||}
\end{equation}
In reality however the transformation of the filament cross-section due to flux expansion and magnetic shear can strongly enhance the curvature drive in regions where the perpendicular scale length of the filament becomes small. The transformation of the $\nabla$ operator along the flux tube is given by $\nabla^{u} = \mathcal{M}\nabla^{d}$ where the transformation matrix $\mathcal{M}$ is given by \cite{MyraPoP2006} 
\begin{equation}
\mathcal{M} = \left(
\begin{array}{c c}
 \left(R_{d}B_{\theta,d}\right)/\left(R_{u}B_{\theta,u}\right) & \left(I_{d} - I_{u}\right)R_{u}R_{d}B_{\theta,u}B_{\theta,d}/B \\ 
 0 & \left(B_{u}R_{u}B_{\theta,u}\right)/\left(B_{d}R_{d}B_{\theta,d}\right) 
 \end{array}
 \right)
 \end{equation}
 where $R$ is the major radius, $B_{\theta}$ is the poloidal magnetic field strength, $B$ is the total magnetic field strength, $I$ is the magnetic shear integrated along the magnetic field line and subscripts $u,d$ refer to upstream (midplane) and downstream values respectively. The diagonal terms represent the decreased gradient in the normal direction and increased gradient in the bi-normal direction as a result of flux-expansion. The off diagonal term represents the increase in the gradient in the normal direction as a result of the shearing of the flux tube. The integration of the curvature drive in the two-region model can now be replaced by
 \begin{equation}
 \int_{0}^{L_{||}} 2c_{s}\rho_{s}\bm{\xi}\cdot\nabla \ln\left(n\right) dL = 2c_{s}\rho_{s} \int_{0}^{L_{||}} \bm{\xi}\cdot\mathcal{M}\nabla^{u}\ln\left(n\right)dL
 \approx \frac{2 c_{s}\rho_{s}}{\delta_{\perp}} L_{eff} 
 \end{equation}
 where 
 \begin{equation}
 L_{eff} = \int_{0}^{L_{||}} 
\left(\begin{array}{c c}\xi_{N} & \xi_{g}\end{array}\right)
\mathcal{M} \left(\begin{array}{c} 1 \\ 1 \end{array}\right) dL 
 \end{equation}
 can be calculated directly from the flux-tube used in the simulations. Here gradient operator has been approximated by 
\begin{equation}
 \nabla^{u} \rightarrow \left(\begin{array}{c}1/ \delta_{\perp}  \\1/ \delta_{\perp}\end{array} \right) 
\end{equation}
 When comparing with the two-region model, $\left|\bm{\xi}\right|L_{||}$ can now be replaced with $L_{eff}$ to provide a consistent comparison. Figure \ref{Fig:2region_2} shows the  scan paths using $L_{eff}$ in $\Lambda$, $\Omega$ space. Also shown is $L_{eff}$ compared to $S$, the distance along the flux tube from the midplane. 
The paths of the $T_{e}$ and $\delta_{\perp}$ scan are both encompassed mainly in the RB regime, in agreement with the velocity and displacement scalings observed in the simulations. The suggests that, despite the complex nature of of the filament motion in 3D, the velocity scaling normal to the flux surface may still be relatively well predicted by the two-region model so long as the effect of the flux-tube transformation on the  curvature drive is accounted for. This comparison therefore supports the use of reduced models like the two-region model as a useful tool for comparison despite its neglect of 3D physics, so long as suitable care is taken in determining $L_{||}$.
 \section{Particle Transport}
\label{Sec:Transp}
\begin{figure}[htbp]
\centering
\includegraphics[width=0.5\textwidth]{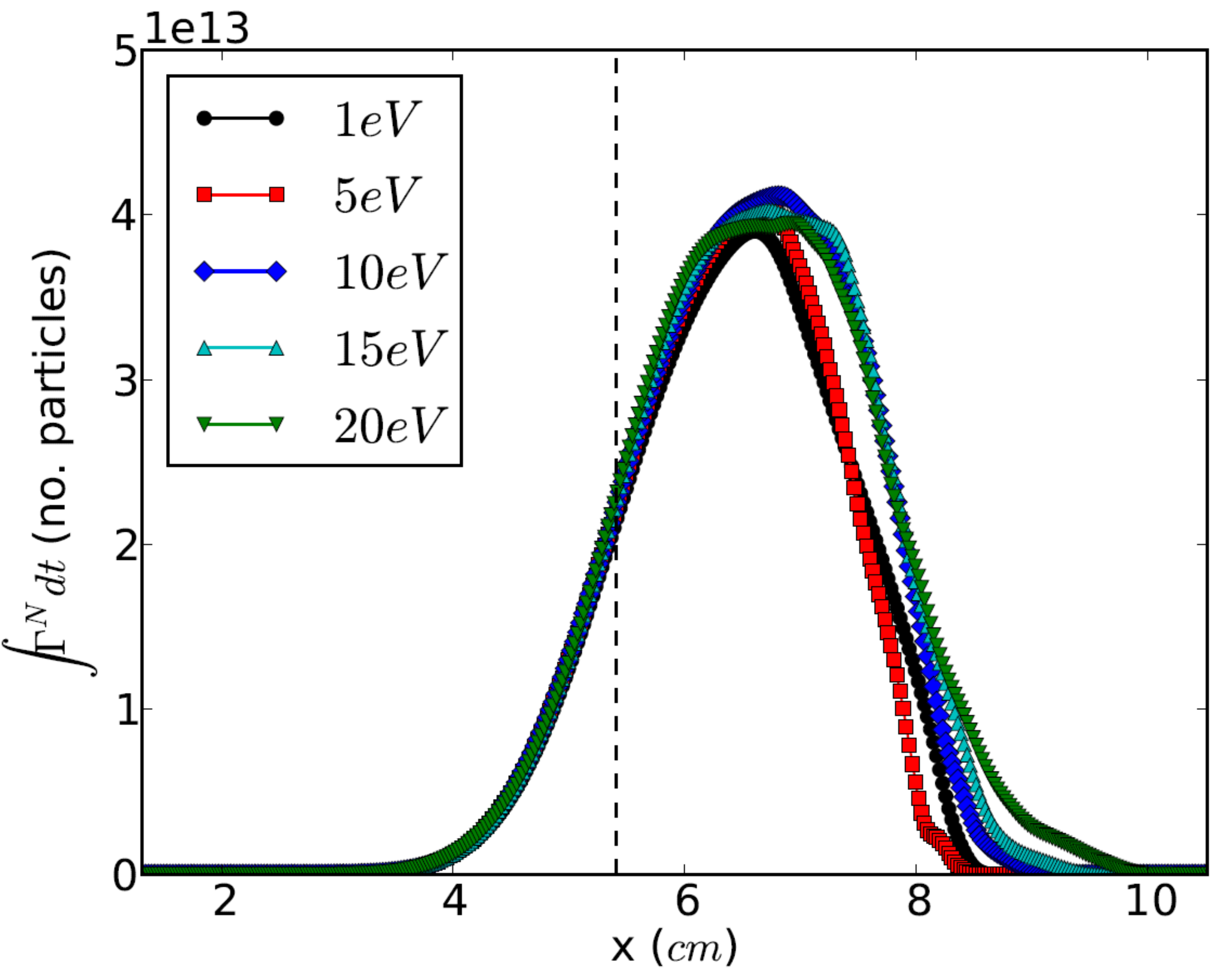}
\caption{Total number of particles crossing each radial location over the simulation time. Little variation is observed between different values of $T_{e}$, though a small fraction of particles reach a further radial extent at $T_{e} = 20$eV. The vertical dashed line indicates the starting position of the filament.}
\label{Fig:Te_Gamma_int}
\end{figure}

\begin{figure}[htbp]
\centering 
\includegraphics[width=0.5\textwidth]{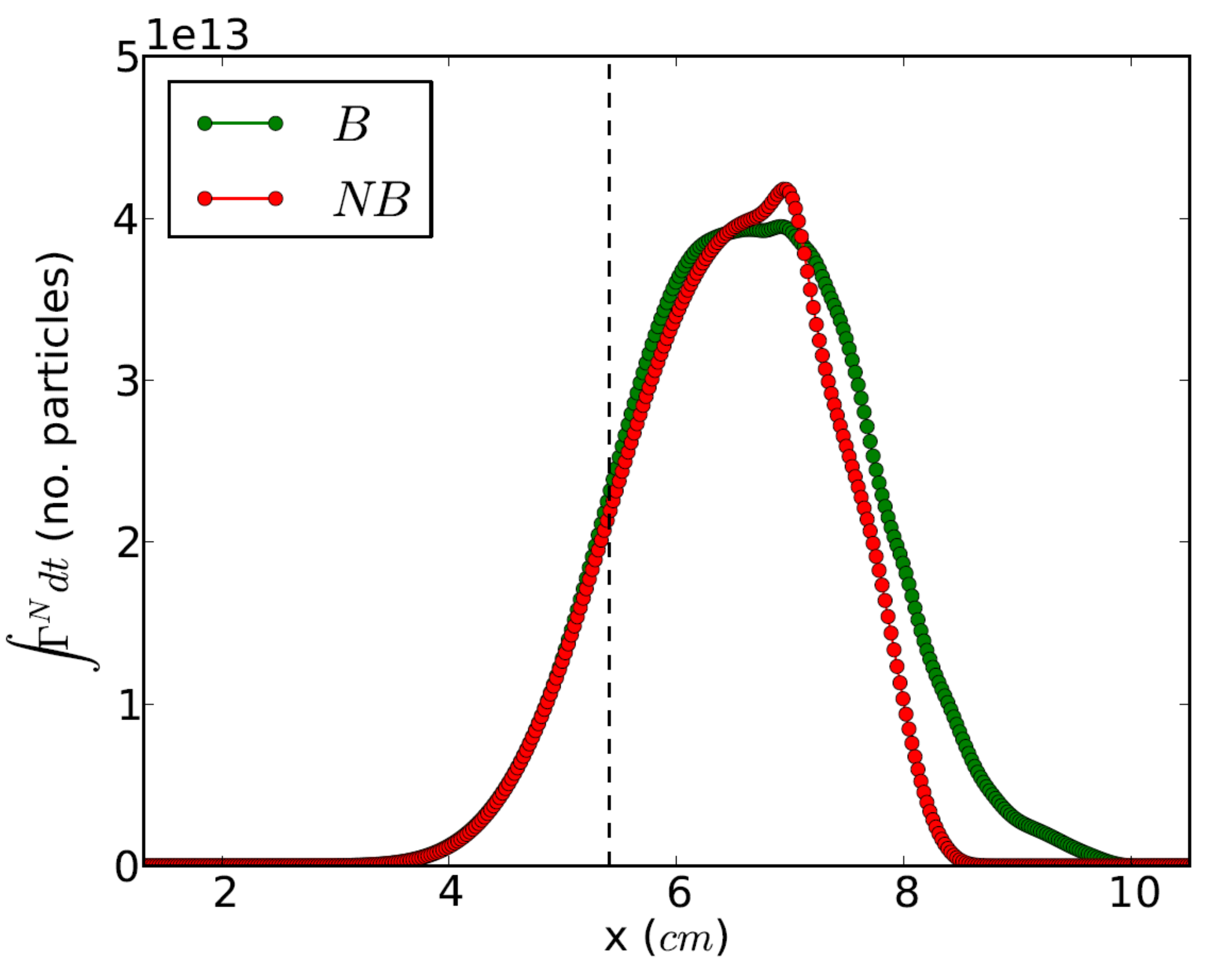}
\caption{Comparison of the total number of particles crossing each radial position with (B) and without (NB) the Boltzmann source included. This indicates that the turbulence induced by the Boltzmann source can enhance particle spreading, though the enhancement is weak.}
\label{Fig:Gama_comp}
\end{figure}

\begin{figure}[htbp]
\centering
\includegraphics[width=0.5\textwidth]{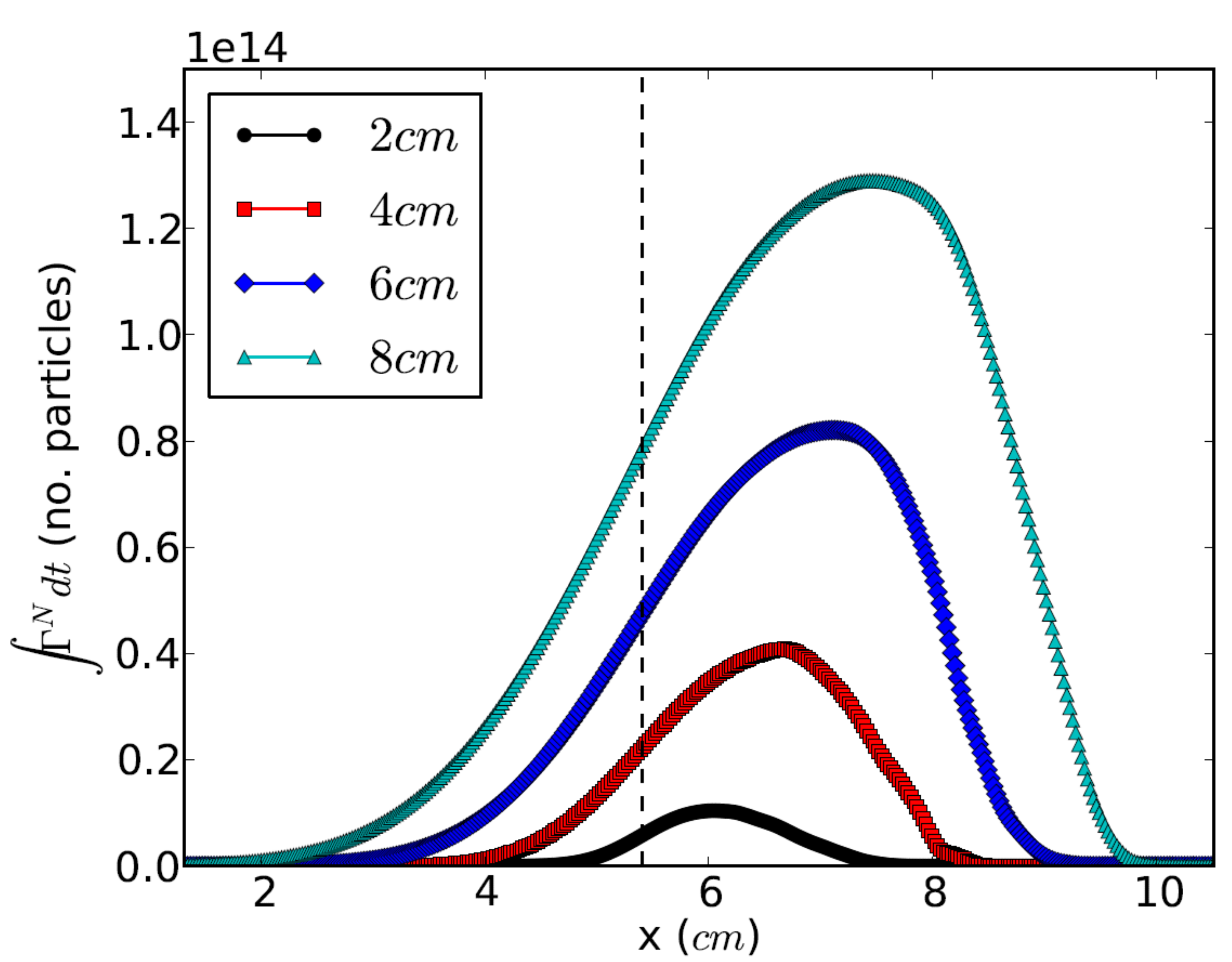}
\caption{Total number of particles crossing each radial location over the $\delta_{\perp}$ scan. A significant increase in the spreading of particles is observed as $\delta_{\perp}$ is increased.}
\label{Fig:d_Gamma_int}
\end{figure}

Analysis of the centre of mass is an effective way of characterising the velocity scaling across parameter scans, but fails to capture the detailed transport of particles due to the evolution of the filament. A more robust analysis of particle transport can be carried out by calculating the particle flux crossing a set of concentric surfaces at each radial location. This provides a radial profile of the instantaneous particle flux due to filament motion, $\Gamma^{N}$, given by
 \begin{equation}
 \Gamma^{N}\left(x,t\right) = \int n_{f}\left(x,t\right)v_{E}^{N}\left(x,t\right)dA\left(x\right)
 \end{equation}
Here $x$ labels the normal coordinate, $n_{f}$ is the filament density, $v_{E}^{N}$ is the $\textbf{E}\times\textbf{B}$ velocity in the normal direction and $dA$ is the surface area of a surface located at $x$. To isolate the dynamics of the filament at the midplane each surface encompasses the entire toroidal domain but extends only $5$mm either side of the midplane in the poloidal direction; as such the surfaces resemble concentric toroidal bands centred on the midplane. By integrating these instantaneous particle fluxes in time it is possible to measure the total number of particles crossing each of the constructed toroidal surfaces during the evolution of the filament. This is shown in figure \ref{Fig:Te_Gamma_int} for filament simulations across the temperature scan.
\\ \\The displacement of particles from their original position increases as the temperature increases. This increase is weak however with the transport of the bulk of the particles remaining relatively constant between simulations. Indeed it is unlikely that variation in particle transport due to filament motion on the scale observed in figure \ref{Fig:Te_Gamma_int} may be observable above experimental uncertainty. This indicates that the role played by electron temperature in determining particle transport mediated by filaments may be weak. Also notable is the fact that only a minority of particles are transported distances $> \delta_{\perp}$. Particle transport into the outer region of the radial domain occurs in higher temperature filaments which are strongly turbulent. To explore whether this turbulence is driving this extra particle transport a second simulation at $20$eV has been run with the Boltzmann source removed thereby removing drift-waves from the system. Comparing the two cases, as shown in figure \ref{Fig:Gama_comp}, it is clear that the particle transport towards the edge of the domain is a result of the drift-wave turbulence. Therefore, whilst drift-waves can act to disperse filaments \cite{AngusPRL2012,AngusPoP2012,EasyPoP2014} they can also enhance particle transport beyond the levels expected in 2D theories.
\\ \\In contrast to the scan in electron temperature, variation of the filament width, $\delta_{\perp}$, can result in a significant change to the particle transport within the filament. This is exhibited in figure \ref{Fig:d_Gamma_int} where the $8$cm filament shows the greatest spreading of particles. 
This is a combined effect of the increased size of the filament containing more particles and of the decrease in the deceleration of the filament with increasing $\delta_{\perp}$ (see figure \ref{Fig:v_dscan}). The magnitude of the different curves in figure \ref{Fig:d_Gamma_int} is largely determined by the number of particles in the initial filament, which is clearly largest for the largest filament. The position of the peak of each curve however is determined by the motion of the filament, since if all filaments move at the same rate, then the surfaces that particles cross remain unchanged. Since the peak in the number of particles crossing radial surfaces is more displaced for larger filaments this indicates that larger filaments are also more efficient at transporting particles into the far SOL. Coupling together the observations that large filaments both contain more particles and transport them more efficiently suggests that large filaments are likely to dominate particle transport into the far SOL.

\section{Discussion and conclusions}
\label{Sec:Conc}
This paper has sought to characterize the transport of particles at the outboard midplane in the SOL,  using a geometry based on the MAST tokamak, due to filament motion through numerical simulation using the BOUT++ code. The presence of parallel density gradients, arising as a result of the magnetic geometry through ballooning of the filament and enhanced diffusion in the divertor region, drives the growth of a component of potential with even parity about the filament centre. This can cause a departure of the filament dynamics from standard 2D behaviour. These effects are most prominent in hotter filaments but begin to have an effect at low temperatures easily obtainable in SOL conditions. The presence of an even component of potential leads to an asymmetric potential dipole in the bi-normal direction causing rotation of the filament density before a break down into drift-wave turbulence occurs. Drift-waves are observed in all filaments across a temperature scan, however in cold filaments they appear as linear instabilities perturbing the filament front but not impacting the motion of the filament significantly. In hotter filaments the drift-waves are non-linearly driven and have a major impact on the filament motion. The large bi-normal displacement may suggest that inter-filament interaction may occur, which cannot be captured by isolated filament simulations and would require significantly more demanding simulations of turbulence.
\\To reconcile the 3D filament simulations with 2D theories the scaling of the centre-of-mass velocity has been investigated across both a scan in temperature and filament width. The velocity of the filament in the direction normal to the flux surface show the characteristics of inertially limited filaments, or to use the terminology of the two region model, are in the inertially limited regime. The simulations were shown to be consistent with the two-region model only when the curvature drive in the divertor region was accounted for. The velocity of the filament in the bi-normal direction showed several features of a scaling based on the Boltzmann response, however the complex dynamics of the filament in this direction stretch the applicability of a velocity scaling argument. It is interesting that the normal and bi-normal velocities appear to show a degree of de-coupling, and that 2D models may still be able to capture the motion in the normal direction relatively well. This is an important point because the most important aspect of particle transport is that which is occurring in the normal direction. If reduced 2D models are capable of capturing the salient features of transport in this direction, despite the neglect of 3D effects which strongly perturb the dynamics of the filament, then they may be a valuable and efficient tool in determining radial particle transport due to filament motion. It is notable that the displacement of the centre of mass of the filaments simulated never exceeds a filament width. In experiment on MAST \cite{DudsonPPCF2008,AyedPPCF2009} filaments are routinely observed to propagate further than their width. This apparent disagreement between simulations and experiment may be rooted in the fact that the model used is a physically reduced system alongside the difficulty in interpreting experimental data. In terms of the model, it is possible that the addition of extra physics into the system will provide more coherent filaments. Finite ion temperature has been shown to provide such an increase in coherence for smaller blobs \cite{MadsenPoP2011} whilst electromagnetic effects have recently been shown to increase the coherence of high $\beta$ filaments \cite{LeePoP2015}. Neither of these effects are included here and may be pursued in the future, though given that filaments in MAST are generally larger than the gyroscale and are relatively low pressure it is unclear how important these two effects may be. Furthermore whilst the centre of mass of the filament in simulations propagates a limited distance the density peak of the filament propagates much further. This presents some ambiguity when comparing the propagation of filaments in experiment with simulations. This work would greatly benefit from a careful validation exercise between simulation and experiment. Such an exercise is already planned for MAST in the near future.
\\It is worth noting that the filament amplitude has not been varied in this study, however in reality will provide another mechanism by which the filament velocity may change. Angus \emph{et.al} \cite{AngusPoP2014-2} have recently demonstrated that for filaments with amplitudes of unitary magnitude, the Boussinesq approximation can break energy conservation which can impact the dynamics of the filament. For large amplitude filaments this effect can be dramatic, however for unit amplitude filaments the deviation is less significant. Furthermore the scalings with respect to temperature and filament width are not likely to change significantly as a result of the Boussinesq approximation. It may be possible however that the cause of the deviation of these simulations from experimental observations on MAST is the amplitude of the simulated filaments. In experiment large amplitude filaments exist, and if it is these filaments that propagate into the far SOL, then they must be modelled. If this is to be achieved then the full system of equations must be considered without the Boussinesq approximation, perhaps in the manner of Angus \emph{et.al} \cite{AngusPoP2014}. 
\\A deeper analysis of particle transport at the outboard midplane has been achieved by calculating the particle flux induced by the motion of the filament through a series of concentric toroidal bands centred on the midplane. This has shown that the presence of drift-wave turbulence in hotter filaments provides a slight enhancement to particle transport by comparison between two simulations, one with and one without the Boltzmann source included (which drives the turbulence). This enhancement is small however and, across the scan in $T_{e}$ the particle transport due to the filament motion remains broadly similar. By contrast the filament width plays a major role in determining the radial transport of particles, with particles in larger filaments being transferred much further in the normal direction. This suggests that a dominant parameter determining the filamentary contribution to radial particle transport is the distribution of filament widths  as opposed to background plasma temperature. 

\section{Acknowledgements}
We would like to thank Drs F. Militello, J. Harrison, M. V. Umansky and J. R. Myra for helpful discussions. We would like to the referees of this paper for their considered comments, which we feel have enhanced the paper. This work was part-funded by the RCUK Energy Programme under grant EP/I501045. This work made use of the HECTor and ARCHER UK National Supercomputing service (http://www.archer.ac.uk) under the Plasma HEC Consortium EPSRC grant number EP/L000237/1. To obtain further information on the data and models underlying this paper please contact PublicationsManager@ccfe.ac.uk. The views and opinions expressed herein do not necessarily reflect those of the European Commission.

\section{References}
\bibliographystyle{prsty}
\bibliography{Bibliography}

\end{document}